\documentclass[12pt]{article}
\def\Slash#1{#1\kern-0.55em\raise.05ex\hbox{/}}
\usepackage{amsmath,amssymb,amsfonts,amscd,cancel,cite,graphicx,caption,subfig}
\begin{document} 


\thispagestyle{empty}
\renewcommand{\thefootnote}{\fnsymbol{footnote}}
\setcounter{footnote}{1}

\vspace*{1.8cm}

\centerline{\Large\bf Resonant Dirac Leptogenesis on Throats}

\vspace*{18mm}

\centerline{\large\bf Andreas Bechinger\footnote{E-mail:
  \texttt{andreas.bechinger@physik.uni-wuerzburg.de}} and
  Gerhart Seidl\footnote{E-mail: \texttt{seidl@physik.uni-wuerzburg.de}}
}

\vspace*{5mm}
\begin{center}
{\em Institut f\"ur Theoretische Physik und Astrophysik,
Universit\"at W\"urzburg,\\
 D-97074 W\"urzburg, Germany}
\end{center}

\vspace*{20mm}

\centerline{\bf Abstract}
\vspace*{2mm}
We consider resonant Dirac leptogenesis in a geometry with three
five-dimensional throats in the flat limit. The baryon asymmetry in
the universe is generated by resonant decays of heavy Kaluza-Klein scalars
that are copies of the standard model Higgs. Discrete exchange symmetries between the throats are responsible for establishing two key features of the model. First, they ensure a near degeneracy of
the scalar masses and thus a resonant decay of the scalars. This
allows for Dirac leptogenesis at low energies close to the TeV
scale. Second, the discrete symmetries connect the observed baryon asymmetry with the Yukawa couplings
of the low-energy theory. As a consequence, we obtain correlations between the low-energy leptonic mixing parameters and the Dirac CP phase that can be tested at future neutrino
oscillation experiments such as neutrino factories.

\renewcommand{\thefootnote}{\arabic{footnote}}
\setcounter{footnote}{0}

\newpage 

\section{Introduction}
One of the major questions in neutrino physics is whether neutrinos
are Dirac or Majorana particles. Currently, considerable experimental
effort is underway \cite{Avignone:2007fu} to measure neutrinoless double beta decay
($0\nu\beta\beta$), which would require the neutrinos to be of
Majorana-type. Neutrino oscillation experiments, however, cannot
distinguish between Dirac and Majorana neutrinos and as long as
$0\nu\beta\beta$ has not been measured, there will always be the
possibility that neutrinos are Dirac particles -- just like all other fermions in the
standard model (SM). 

One advantage of having Majorana neutrinos is that the smallness of the
observed light neutrino masses $\sim 10^{-1}\,\text{eV}$
could be understood in terms of the seesaw mechanism
\cite{typeIseesaw,typeIIseesaw} which establishes a connection to grand unified theories (GUTs). The type-I seesaw mechanism \cite{typeIseesaw} offers, furthermore, the possibility to understand the observed  baryon
asymmetry in the universe (BAU) \cite{Sakharov,WMAP} through baryogenesis via
leptogenesis \cite{Fukugita:1986hr}. In standard leptogenesis, the BAU is produced by the decay of the heavy SM singlet Majorana neutrinos that generate small neutrino masses via the seesaw mechanism (for reviews see,
e.g., \cite{Buchmuller} and \cite{Riotto:1998bt}). If, however, the neutrinos are Dirac particles, the original leptogenesis scenario would  no longer apply since the SM singlet neutrinos would have zero Majorana mass. The BAU can then, instead, be generated by Dirac leptogenesis
\cite{Dick:1999je}. Several studies have shown that Dirac leptogenesis may indeed be responsible for the BAU \cite{Murayama:2002je,Thomas:2005rs,Gu:2006dc,Gherghetta:2007au}.

In Dirac leptogenesis, the BAU is generated by the decay of heavy copies of the SM Higgs doublet into leptons. Due to the smallness of
the Dirac neutrino Yukawa couplings the resulting lepton asymmetry can be stored sufficiently long in the singlet neutrino sector to allow for successful baryogenesis via
sphaleron processes \cite{sphalerons}. The original version of Dirac leptogenesis,
however, raises a couple of questions. First, the scenario does not
address the origin of the heavy copies of the SM Higgs. Second, Dirac
leptogenesis would serve as a GUT-scale leptogenesis scenario \cite{GUTscale} by preferably taking place at rather large energies near $\sim 10^{16}\,\text{GeV}$. This may, however, get into
conflict with standard inflationary models and the gravitino
problem \cite{Davidson:2002qv}. Besides that, the Yukawa couplings
responsible for leptogenesis seem to be completely unrelated to the
Yukawa couplings giving rise to the observed neutrino masses. In other
words, in the original scenario for Dirac leptogenesis, a measurement
of the low-energy lepton mass and mixing parameters would have no
connection with the BAU.

In this paper, we consider a model for Dirac leptogenesis that addresses all of these problems. The model makes use of discrete symmetries to (i) implement resonant leptogenesis at low energies close to the scale of electroweak symmetry breaking (EWSB)  \cite{resonant,resonantreview} (see also \cite{resonantstudies}) and to (ii) relate the observed BAU with the low-energy lepton mixing parameters measurable in neutrino oscillation experiments. For this purpose, we work in the
flat limit of a five-dimensional (5D) background with several ``throats'' that can emerge
from flux compactification in string theory
\cite{throats,throatapplications}.  The field theory on this multi-throat background \cite{Cacciapaglia:2006tg,Agashe:2007jb} allows to identify the heavy scalars necessary for Dirac leptogenesis with Kaluza-Klein (KK)
excitations and use the field separation on the throats to naturally implement nearly mass degenerate scalars that decay resonantly.

The paper is organized as follows: In
Sec.~\ref{sec:Diracleptogenesis}, we briefly review the idea of Dirac
leptogenesis. Next, in Sec.~\ref{sec:Model}, we present our model for
Dirac leptogenesis on a background with three throats. In Sec.~\ref{sec:BCs}, we discuss
the boundary conditions of the 5D scalars and fermions along with the resulting wavefunction profiles and the Yukawa couplings. Then, in
Sec.~\ref{sec:Leptonasymmetry}, we determine the range of Yukawa
couplings necessary for successful Dirac leptogenesis and discuss the connection of the BAU with the low-energy lepton mixing parameters. In Sec.~\ref{sec:Summary}, we present our summary and conclusions. Finally, in the appendix, we give further examples for correlations of the low-energy mixing parameters.

\section{Brief Review of Dirac Leptogenesis}\label{sec:Diracleptogenesis}
Let us start by giving a short review of the Dirac leptogenesis scenario
 proposed in \cite{Dick:1999je}.  We assume the SM gauge group $G_{SM}=SU(3)_C\times SU(2)_L\times U(1)_Y$.
 Unless otherwise stated, it is, in the following, understood that a left-handed (LH) fermion is in an $SU(2)_L$ doublet representation while a right-handed (RH)  fermion is an $SU(2)_L$ singlet. We denote the LH lepton doublets of the SM by $\ell_{La}$ and
 the RH charged leptons by $e_{Ra}$, where $a=1,2,3$ is the
 generation index. Moreover, we extend the SM by three RH neutrinos $\nu_{Ra}$ which are total gauge singlets of the SM gauge group.

\begin{figure}[t!]
\begin{center}
\includegraphics*[width=0.7\textwidth]{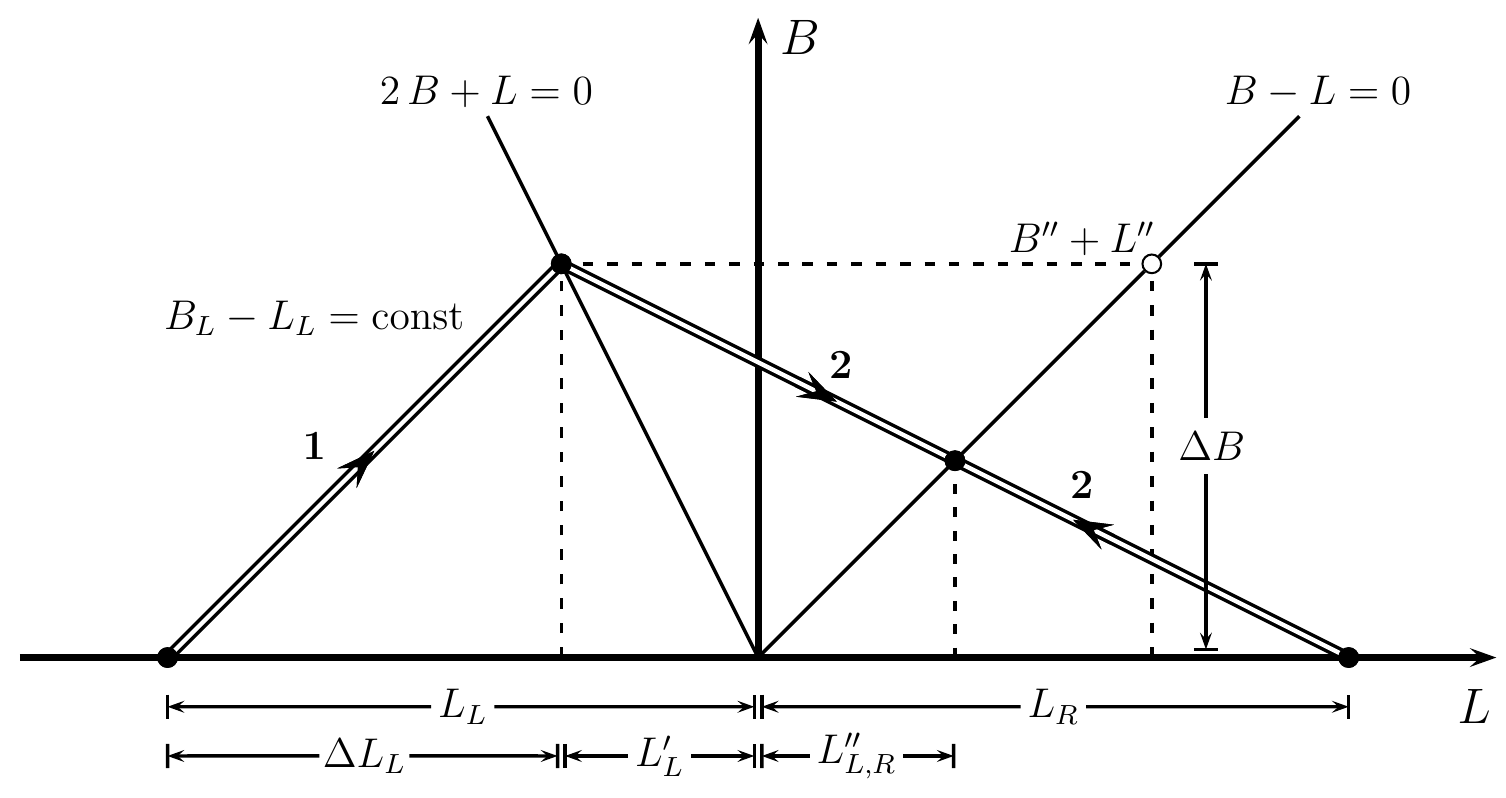}
\caption{Generation of baryon asymmetry in Dirac
  leptogenesis \cite{Dick:1999je}. Starting from LH and RH lepton asymmetries
  $L_R=-L_L>0$ in the neutrino sector, sphaleron processes (1) produce
  a baryon asymmetry $\Delta B=\Delta L_L$ from $L_L$. For sufficiently
  small Dirac neutrino Yukawa couplings, LR equilibration processes (2) set in later and
  lead to a final total lepton asymmetry $L''=L_L''+L_R''=B''=\Delta B$.}\label{fig:Diracleptogenesis}
\end{center}
\end{figure}
\begin{figure}
\begin{center}
\includegraphics*[width=0.50\textwidth]{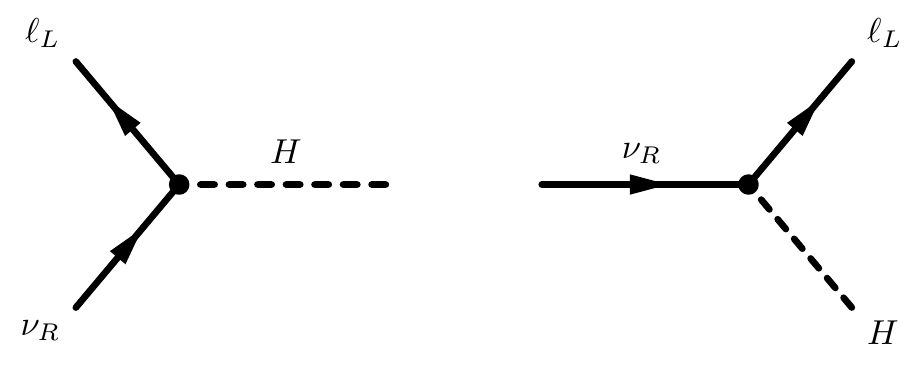}
\caption{LR equilibration processes
  $\bar{\ell}_L+\nu_R\rightarrow \bar H$ and $\nu_R\rightarrow \ell_L+\bar H$ changing $L_L$ and $L_R$.}\label{fig:LRequilibration}
\end{center}
\end{figure}

Assume now that the early universe is baryon and lepton symmetric, i.e. the total baryon number $B$ as well as the total lepton number $L$ both vanish. Furthermore, there shall be no asymmetry between the
LH and RH sectors. In other words, the baryon number $B=B_L+B_R$ and the lepton number
$L=L_L+L_R$ are both zero and vanish also in the LH and RH sectors, separately. Consider next the case where
some process has produced a relative asymmetry $L_R=-L_L>0$ in the neutrino sector,
i.e. $L_L$ and $L_R$ come from an excess of $\bar{\ell}_{La}$ and $\nu_{Ra}$, respectively.
The LH and RH baryon
and lepton numbers will be affected by two types of processes: (1) sphaleronic vacuum to vacuum transitions \cite{sphalerons} and (2)
left-right (LR) equilibration processes (see Fig.~\ref{fig:Diracleptogenesis}).
Note that $SU(2)_L$ sphaleronic processes only act on the LH sector. They violate $B_L$ and $L_L$ by 3 units each, i.e.~they are conserving $B_L-L_L$ but not $B_L+L_L$. LR equilibration
processes, by contrast, conserve $B$ and $L$ separately but violate $B_{L,R}$ and
$L_{L,R}$. Suppressing generation indices, the Yukawa couplings between $\bar{\ell}_L$ 
and $\nu_R$ to the SM Higgs
field $H$ give rise to LR equilibration processes such as $\bar{\ell}_L+\nu_R\rightarrow H$ and $\nu_R\rightarrow\ell_L+H$ (see Fig.~\ref{fig:LRequilibration}).
For sufficiently small Dirac neutrino Yukawa couplings,
sphaleron processes will dominate the LR equilibration processes in
the neutrino sector and $L_L$ will be partly converted into a nonzero
LH baryon asymmetry $\Delta B_L>0$. However, LR equilibration
processes in the baryonic sector, which also include $SU(3)_c$ sphaleronic
processes \cite{Mohapatra:1991bz}, are fast and transfer half of
$\Delta B_L$ into a RH baryon number $\Delta B_R>0$. Consequently,
the $SU(2)_L$ sphaleronic processes reach equilibrium at
$2 B_L+L_L=0$. This changes $L_L$ by an
amount $\Delta L_L=2\Delta B_L$ to a new lepton asymmetry
$L_L'=L_L+\Delta L_L$. Since $L_R$ remains
unaffected by sphalerons ($L_R'=L_R$), the subsequent LR equilibration
processes convert $L_L'$ and $L_R$ into the final lepton asymmetries
$L_L''$ and $L_R''$, which are equal and given by $L_{L,R}''=\Delta
L_L/2>0$. Thus, we arrive at a final total positive lepton asymmetry
$L''$ that is $L''=L_L''+L_R''=\Delta L_L>0$. At the same time, the
final total baryon asymmetry is $B''=2\Delta B_L=\Delta B$. As already mentioned,
this requires the neutrino Yukawa couplings to be sufficiently small so as to allow the LR equilibration to take place only after sphaleron processes have dropped out of thermal equilibrium. A numerical study shows that this becomes possible for neutrino Yukawa couplings of the order $\lesssim 10^{-8}$ \cite{Dick:1999je}. Therefore, if the neutrinos were Majorana particles, with the observed small neutrino mass scale $m_\nu\sim 10^{-1}\,\text{eV}$ generated by the type-I seesaw mechanism, the neutrino Yukawa couplings to the SM Higgs field would be $\sim 1$, which is by many orders too large. If, instead, the neutrinos are Dirac particles, the neutrino
Yukawa couplings will be of the order $m_\nu/v\sim 10^{-12}$, where $v\sim 10^2\;\text{GeV}$ is the SM Higgs vacuum expectation value (VEV), and sphaleron
processes will dominate LR equilibration in the neutrino sector as required for successful Dirac
leptogenesis.

Let us briefly compare with the case of the SM (without massive neutrinos). In the SM, all Yukawa
couplings in the quark and lepton sectors are $\gg 10^{-8}$ and LR equilibration would be taking place roughly at the same time as the sphaleron
processes. Starting with arbitrary $B$ and $L$, the LR equilibration would then quickly drive
$B+L$ to values with $B-L=0$ such that sphaleron processes can only give
$B+L=0$ and, thus, yield zero net baron and lepton asymmetries. To make the
above mechanism for leptogenesis work, we thus require Dirac neutrinos which can provide sufficiently small neutrino Yukawa couplings.

\section{Throat Geometry}\label{sec:Model}
\begin{figure}[t]
\begin{center}
\includegraphics*[width=0.5\textwidth]{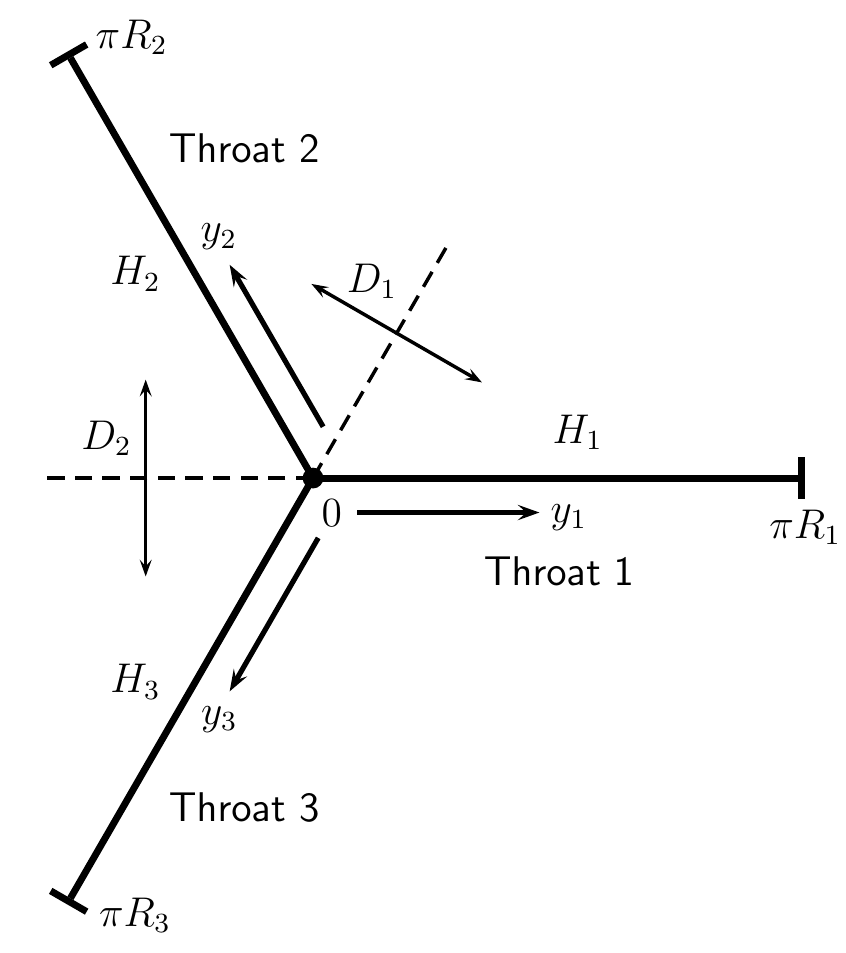}
\caption{Geometry of the three throats. The throats are described by
  intervals with coordinates $y_1,y_2,$ and $y_3$, that intersect at the
UV brane at $y_1=y_2=y_3=0$. The Higgs doublets $H_1$, $H_2$,
and $H_3$, are living separately on the 1st, 2nd, and 3rd throat,
respectively. All the other fields, such as the SM fermions, are
  propagating on all three throats. The throats are related by exchange symmetries $D_1$ and $D_2$.}\label{fig:throats}
\end{center}
\end{figure}

We will now be concerned with a model for Dirac neutrinos that
generates the observed baryon asymmetry via Dirac leptogenesis as
discussed in Sec.~\ref{sec:Diracleptogenesis}. Consider for this
purpose three intervals in 5D flat space which are glued together at a single point as shown in
Fig.~\ref{fig:throats}. We will call the intervals throats. The coordinates on the three throats are respectively $z_1^M=(x^\mu,y_1),
z_2^M=(x^\mu,y_2),$ and $z_3^M=(x^\mu,y_3)$, where the 5D Lorentz indices are denoted by capital Roman letters $M=0,1,2,3,5$, while
the usual 4D Lorentz indices are symbolized by Greek letters
$\mu=0,1,2,3$. The coordinates $y_1,y_2,$ and $y_3$, describe the 5th
dimension for the three throats. The physical space is thus defined by $0\leq y_1\leq\pi R_1,0\leq y_2\leq\pi R_2,$ and
$0\leq y_3\leq\pi R_3,$ where $R_1,R_2,$ and $R_3,$ denote the size of
the throats. The intersection point at
$y_1=y_2=y_3=0$ will be called the UV brane and the
 endpoints of the intervals at $y_1=\pi R_1,y_2=\pi R_2,$ and $y_3=\pi
 R_3$, will be denoted as IR branes. It will be useful to
characterize this throat geometry by reflection symmetries interchanging the 1st and 2nd
as well as the 2nd and 3rd throat (see Fig.~\ref{fig:throats}). We
will comment on these symmetries and how they are broken later.

On the three throats, we assume the SM gauge group $G_\text{SM}$. The scalar sector contains three
5D Higgs doublet fields $H_1,H_2,$ and $H_3$, that carry the same $G_\text{SM}$ quantum numbers as the usual SM Higgs field. We assume that the three Higgs doublets live on separate
throats: Each of the fields $H_1,H_2,$ and $H_3$, is propagating on the first ($H_1$), second ($H_2$), and third ($H_3$) throat, respectively.
By separating the scalar fields on the throats, we can, in the following, neglect the mixing between the scalars. The action of the Higgs doublets is then
\begin{equation}
\mathcal{S}_H=\int d^4x\sum_{i=1}^3\int_0^{\pi R_i}dy_i\,
\mathcal{L}_{H_i},\label{eq:Sskalar}
\end{equation}
where the 5D scalar Lagrangian density is
\begin{equation}
\mathcal{L}_{H_i}=(D_MH_i)^\dagger
 D^MH_i-\mu_i^2\,\,H_i^\dagger
 H_i+\lambda_i(H_i^\dagger H_i)^2,\label{eq:Lscalar}
\end{equation}
in which $D_M$ is the covariant derivative. We will later discuss how
the SM gauge group is spontaneously broken when only $H_1$ acquires a
non-zero VEV in the bulk of the 1st throat.

Let us now focus on the lepton sector only (the discussion for quarks should be along the same lines). We suppose that the SM leptons propagate on all three throats. 
For this purpose, we start on each throat with 5D lepton fields
$L_{i,a},E_{i,a},$ and $N_{i,a}$, where $i=1,2,3$ labels the throat
and $a=1,2,3$ is the generation index. Each fermion with label $i$
propagates, like $H_i$, only on the throat $i$. The fields $L_{i,a}$ carry
the quantum numbers of the LH SM lepton doublets, $E_{i,a}$ of the
RH SM charged leptons, while $N_{i,a}$ are SM singlet
neutrinos. As the 5D action of the fermions we take
\begin{equation}\label{eq:S5DPsi}
\mathcal{S}_\Psi^\text{5D}=\sum_{\Psi=L,E,N}
\int d^4x\sum_{i,a=1}^{3}\int_0^{\pi R_i}dy_i\,\Big(\frac{\text{i}}{2}\bar \Psi_{i,a}\Gamma^M\overleftrightarrow{\partial_M}\Psi_{i,a}-m^\Psi_{i,a}\bar \Psi_{i,a}\Psi_{i,a}\Big),
\end{equation}
where $\Psi=L,E,N,$ denotes the fermion species, $m^\Psi_{i,a}$ are the bulk masses of the $\Psi_{i,a}$,
$\overleftrightarrow{\partial_M}=\overrightarrow{\partial_M}-\overleftarrow{\partial_M}$,
$\Gamma^\mu=\sigma^1\otimes\sigma^\mu$,
  $\Gamma^5=\text{diag}(\text{i}\mathbf{1}_2,-\text{i}\mathbf{1}_2)$, $\sigma^0=-\mathbf{1}_2$, and
  $\sigma^i$ ($i=1,2,3$) are the Pauli matrices. Note that the fields $L_{i,a},E_{i,a},$ and $N_{i,a}$, are
vector-like in 5D. By imposing appropriate boundary conditions and interactions at
the UV brane, we will later show how to obtain from these fields chiral
fermion zero modes that can be identified with the SM fermions.

We suppose that the 5D fermions couple to the scalar doublets only via Yukawa
interaction terms localized at the IR branes of the throats:
\begin{equation}\label{eq:SY}
\mathcal{S}_Y=\int d^4x\sum_{i=1}^3\int_0^{\pi
   R_i}dy_i\,\mathcal{L}^5_{Yi}+\text{h.c.},
\end{equation}
in which the 5D Yukawa coupling Lagrangians $\mathcal{L}_{Yi}^5$ are
\begin{equation}\label{eq:L5DYi}
 \mathcal{L}^5_{Yi}=\sum_{a,b=1}^3\delta(y_i-\pi
 R_i)[(Y^5_i)_{ab}H_i\bar{L}_{i,a}N_{i,b}+(\widetilde{Y}^5_i)_{ab}\widetilde{H}_i\bar{L}_{i,a}E_{i,b}]+\text{h.c.},
\end{equation}
where $\widetilde{H}_i=\text{i}\sigma_2H_i^\ast$, while $Y^5_i$ and
$\widetilde{Y}_i^5$ are the complex $3\times 3$ lepton Yukawa coupling
matrices for $H_i$ and $\widetilde{H}_i$ in the 5D theory.

We assume that the throats are subject to exchange symmetries (see Fig.~\ref{fig:throats}) which act on the scalar doublets $H_i$, the fermions $\Psi_{i,a}$, and the throat coordinates as
\begin{eqnarray}\label{eq:D1}
D_1&:&H_1\leftrightarrow H_2,\quad
\Psi_{1,a}\leftrightarrow
\Psi_{2,a},\quad
y_1\leftrightarrow y_2,
\end{eqnarray}
and
\begin{eqnarray}\label{eq:D2}
D_2&:&H_2\leftrightarrow H_3,\quad
\Psi_{2,a}\leftrightarrow P^\Psi_{ab}\Psi_{3,b},
\quad y_2\leftrightarrow y_3.
\end{eqnarray}
Here, $(P_{ab}^\Psi)=P^\Psi$ are
$3\times 3$ matrix representations of some discrete flavor
symmetry, i.e.~the $P^\Psi_{ab}$ act on the generation indices. As a simple example, we will consider
\begin{equation}\label{eq:Pij}
P^L=
\left(\begin{matrix}
0 & -1 & 0\\
1 & 0 & 0\\
0 & 0 & -1
\end{matrix}
\right),
\quad
P^E=P^N=\mathbf{1}_3,
\end{equation}
but other choices, such as those presented in the appendix, are also possible. The matrix $P^L$ yields a representation of an element of the group $\Delta(24)\sim(Z_2\times Z_2)\rtimes S_3$ (from the class $6\,C_3^{(1)}$) \cite{Escobar:2008vc} and generates a $Z_4$ subgroup of $\Delta(24)$. 
In Fig.~\ref{fig:throats}, the symmetry $D_1$ ($D_2$) corresponds to a
reflection symmetry with respect to the dashed line between
the throats 1 and 2 (2 and 3). Note that the symmetries $D_1$ and $D_2$ require
the throats to have equal lengths $R_1=R_2=R_3$. Moreover, for the
choice of matrices in (\ref{eq:Pij}), the symmetries $D_1$ and
$D_2$ establish among the Yukawa coupling matrices the identities
\begin{equation}\label{eq:Yukawarelations}
Y^5_1=Y^5_2=P^L Y^5_3,\quad \widetilde{Y}^5_1=\widetilde{Y}^5_2= P^L \widetilde{Y}^5_3.
\end{equation}
As we will see below, to obtain a
realistic light fermion spectrum, we need to break $D_1$ and $D_2$ at the UV brane.

Let us briefly comment on how the symmetry $D_2$ could emerge from the
product of two finite groups via spontaneous symmetry breaking. We
begin with a group $D_2'\sim Z_2\times G$, where $G$ is the flavor symmetry group that acts on the fields on the third throat and is generated by the matrices in
(\ref{eq:Pij}). The group $Z_2$ is given by the exchange symmetry $Z_2:\,X_{2}\leftrightarrow X_3,\,y_2\leftrightarrow
y_3$, where $X=H,L,E,N$ (generation indices have been neglected). Assume now a SM singlet scalar field $S$ that
carries a charge $+1$ under the $Z_2$ symmetry and transforms under
application of the flavor symmetry transformation in
(\ref{eq:Pij}) as $S\rightarrow -S$. When $S$
acquires a nonzero VEV, $D_2'$ will be broken at
some high scale down to the
subgroup $D_2$ of (\ref{eq:D2}). However, instead of expanding further on the details
of the possible underlying symmetry groups at high energies, we will, in the
following, only be concerned with the symmetries $D_1$ and $D_2$ of the low-scale theory.

\section{Wavefunction Profiles}\label{sec:BCs}
In this section, we consider the boundary conditions for the 5D
scalar and fermion fields, determine the mass
spectra and wavefunctions of the bulk scalars, and demonstrate the
exponential wavefunction localization of the fermion zero modes on the throats.

\subsection{Scalar Boundary Conditions}
The scalar doublets on the three throats are supposed to be subject to the following BCs
\begin{subequations}
\begin{eqnarray}\label{eq:scalarBCs}
\text{at the IR branes}&:&\left.\partial_{y_i}H_i\right|_{y_i=\pi R_i}=0,\\
\text{at the UV brane}&:&\left.\partial_{y_1}H_1\right|_{y_1=0}=0,\quad
\left.H_{2,3}\right|_{y_{2,3}=0}=0.
\end{eqnarray}
\end{subequations}
Note that we have on the first throat Neumann BCs
at both endpoints for the field $H_1$, whereas the fields $H_{2,3}$ have Neumann
BCs at the UV brane and Dirichlet BCs at the IR branes. The most
general flat space KK expansion of the scalars, consistent with the
BCs in (\ref{eq:scalarBCs}) is for $H_1$ given by
\begin{equation}\label{eq:KKH1}
 H_1(x_\mu,y_1)=\frac{1}{\sqrt{\pi
     R_1}}\Big[H_1^{(0)}(x_\mu)+\sqrt{2}\sum_{n=1}^{\infty}H_1^{(n)}(x_\mu)\,\text{cos}\Big(\frac{ny_1}{R_1}\Big)\Big],
\end{equation}
while the KK expansions for the fields $H_{2,3}$ read
\begin{equation}
 H_i(x_\mu,y_i)=\sqrt{\frac{2}{\pi R_i}}\sum_{n=1}^{\infty}H_i^{(n)}(x_\mu)\,\text{sin}\Big(\frac{(2n-1)y_i}{2R_i}\Big),
\end{equation}
where $i=2,3$. Note the important fact that the Dirichlet BCs at the UV brane have projected out the zero
modes of $H_{2,3}$, such that only $H_1$ will have a zero
mode. At the same time, the Neumann BCs at the IR branes ensure that the
$H_i$ are non-vanishing there. $D_1$ is broken by the different BCs
for $H_1$ and $H_2$ at the UV brane. Moreover, as we will explain
further below, $D_1$ is broken by the bulk mass terms for the SM
singlet neutrinos $N_{i,a}$. Different from the symmetry $D_1$,
however, $D_2$ remains almost completely intact in the scalar sector.

Denoting by $M_n(H_i)$ the mass of the $n$th KK state
$H_i^{(n)}$ of $H_i$ at zero temperature, we thus arrive for
$H_1$ and $H_{2,3}$ at the mass squares of the KK states
\begin{equation}
 M^2_n(H_1)=\Big[-\mu_1^{2}+\Big(\frac{n}{\pi R_1}\Big)^2\Big],\quad
 M^2_n(H_i)=\Big[-\mu_i^2+\Big(\frac{2n-1}{2\pi R_i}\Big)^2\Big]\quad(i=2,3),
\end{equation}
where $n=0,1,\dots$ for $H_1$ and $n=1,2,\dots$ for $H_2$ and $H_3$. Notice that the discrete symmetry $D_2$ in (\ref{eq:D2}) establishes
$\mu_2^2=\mu_3^2$ as well as $R_2=R_3$. The mass squares $M_n^2(H_2)$
and $M_n^2(H_3)$ will therefore, up to small corrections, be practically degenerate. As we will see later, for successful Dirac leptogenesis, we
will need small mass-squared splittings of the order
\begin{equation}\label{eq:splitting}
|M^2_n(H_2)-M^2_n(H_3)|/M^2_n(H_2)\sim 10^{-8}.
\end{equation}
Such small relative mass-squared splittings may be induced, e.g., at the quantum level, but we will not further specify an origin of this splitting here. The potential of the 5D scalar doublet $H_1$ with the KK expansion given in
(\ref{eq:KKH1}) has a local minimum for the VEVs \cite{Muck:2001yv}
\begin{equation}\label{eq:H1VEVs}
\langle H_1^{(0)}\rangle=\frac{1}{\sqrt{2}}\left(
\begin{matrix}
 0\\
 v
\end{matrix}
\right),\quad
 \langle H_1^{(n)}\rangle=0,
\end{equation}
where $v$ is a real parameter with mass dimension $+1$ and
$n=1,2,\dots$ In other words, only the zero mode $H_1^{(0)}$ acquires
a non-zero VEV while all higher KK excitations have zero
VEVs. Qualitatively, this is because only the zero mode has a
negative mass square (coming from the potential), while the higher KK
excitations have, for a sufficiently large
compactification scale, always positive mass-squares. Similarly, since
the zero modes of $H_{2,3}$ have been projected out by the
Dirichlet BCs, we will take for $H_{2,3}^{(n)}$ $(n\geq 1)$
the VEVs
\begin{equation}\label{eq:VEVs}
 \langle H_{2,3}^{(n)}\rangle=0,
\end{equation}
i.e. the VEVs of all KK excitations of $H_{2,3}$
vanish. Therefore, only $H_1^{(0)}$ with the VEV given in (\ref{eq:H1VEVs}) will be responsible for EWSB and for generating masses for the SM fermions from the Yukawa interactions in (\ref{eq:L5DYi}).

\subsection{Fermion Boundary Conditions}
Since fermions in 5D are vector-like, we have to
impose appropriate BCs in order to obtain a chiral 4D
theory. In doing so, we will apply the techniques introduced in
\cite{Cacciapaglia:2006tg,Csaki:2003} for multi-throat geometries. For this purpose, we write, neglecting
generation indices, the 5D fermions on the $i$th throat $\Psi_i$ ($\Psi=L,E,N$) as Dirac spinors of the form
\begin{equation}
\Psi_i=
\left(
\begin{matrix}
\Psi_{Li}\\
\Psi_{Ri}
\end{matrix}
\right),
\end{equation}
where $\Psi_{Li}$ and $\Psi_{Ri}$ denote two-component Weyl spinors and $i=1,2,3$ labels the throat on which $\Psi_i$ lives. The 5D action of
the Dirac spinors is given by $\mathcal{S}_\Psi^\text{5D}$ in
(\ref{eq:S5DPsi}). In absence of brane-localized operators, the
equations of motion for $\Psi_i$ read
\begin{subequations}\label{eq:EoM}
\begin{eqnarray}
-\text{i}\bar{\sigma}^\mu\partial_\mu\Psi_{Li}-\partial_5\Psi_{Ri}+m^\Psi_i\Psi_{Ri}&=&0,\label{eq:EoML}\\
-\text{i}\sigma^\mu\partial_\mu\Psi_{Ri}+\partial_5\Psi_{Li}+m^\Psi_i\Psi_{Li}&=&0.
\end{eqnarray}
\end{subequations}
Consider now for $\Psi_{Ri}$ at both endpoints of the $i$th throat Dirichlet BCs 
$\Psi_{Ri}|_{0,\pi R_i}=0$, which lead for
$\Psi_{Li}$ to the appearance of a single chiral zero mode with an
exponential 5D wavefunction $\sim \text{exp}(-m^\Psi_iy_i)$ propagating on
all three throats (for a discussion of exponential localizations of wavefunctions see \cite{Jackiw:1975fn,Kaplan:2001}). This is achieved by connecting the throats by a brane-localized action
$\mathcal{S}_\text{UV}$ at the UV brane. This action couples the $\Psi_{Li}$ to some extra fields
$\xi^\Psi_{Ri}$ ($i=1,2$) which are localized at the UV brane:
\begin{equation}
\mathcal{S}_\text{UV}=\int d^4x\sum_{i=1}^{3}\sum_{j=1}^2\int_0^{\pi
  R_i}dy_i\,\big((m_\text{UV}^\Psi)^\frac{1}{2}\mathcal{K}^\Psi_{ij}\bar{\Psi}_{Li}\xi^\Psi_{Rj}+\text{h.c.}\big)\,\delta(y_i),
\end{equation}
where $m_\text{UV}^\Psi$ is a UV brane mass parameter and $\mathcal{K}^\Psi_{ij}$
is a dimensionless $3\times 2$ rank 2 matrix. Note that since the equations
of motion in (\ref{eq:EoM}) are first order differential equations, the brane-localized
operators will lead to a discontinuity of the wavefunction of the RH
field $\Psi_{Ri}$ at $y_i=0$, i.e.~$\text{lim}_{\epsilon\rightarrow 0}
\Psi_{Li}(x^\mu,y_i=\epsilon)\neq\Psi_{Ri}(x^\mu,y=0)=0$. Including the brane
interaction term
$(m_\text{UV}^\Psi)^\frac{1}{2}\mathcal{K}_{ij}^\Psi\xi^\Psi_{Rj}\delta(y_i)$ in
(\ref{eq:EoML}), we then obtain for the LH fields the BCs \cite{Cacciapaglia:2006tg}
\begin{equation}\label{eq:fermionBCs}
\sum_{i=1}^3\mathcal{K}^\Psi_{ij}\bar{\Psi}_{Li}|_{y_i=0}=0.
\end{equation}
As a consequence, two of the three zero modes decouple
for large $m_\text{UV}^\Psi$, leaving the remaining zero mode as a single
chiral field propagating on all three throats. The
wavefunction of this mode is
\begin{equation}\label{eq:fermionzeromode}
\Psi_{L}^{(0)}(x,y_i)=A^\Psi_0\,
\text{exp}(-m^\Psi_iy_i)\,\psi^0_{L}(x)\quad\text{with}\quad
A^\Psi_0=\bigg[\sum_{i=1}^3\frac{1-\text{exp}(-2m^\Psi_i\pi R_i)}{2m^\Psi_i}\bigg]^{-\frac{1}{2}},
\end{equation}
where $\psi^0_L(x)$ is a 4D Weyl spinor with mass dimension 3/2. We can see from (\ref{eq:fermionzeromode}) that the actual localization of the zero modes depends on the signs of the bulk masses $m^\Psi_i$, which can be
positive or negative. We thus obtain the following possible localizations: For
$m^\Psi_1=m^\Psi_2=m^\Psi_3<0$, the zero mode $\Psi_L^{(0)}$ is localized at the three IR branes, for
$m^\Psi_1=m^\Psi_2=m^\Psi_3>0$, it is localized at the UV brane, and
for only one or two $m^\Psi_i<0$, it is localized at the IR branes of
the throats with positive $m_i^\Psi$. In a similar way, we can localize RH fermion zero modes on
the UV and IR branes of the throats by replacing in the above considerations
the LH and RH fields.

The brane-localized Yukawa couplings at the IR branes in
(\ref{eq:L5DYi}) lead  to discontinuities of the wavefunctions of
$L_{Ri},E_{Li},$ and $N_{Li}$, at $y_i=\pi R_i$. The
wavefunctions obey at the IR branes the BCs
\begin{subequations}
\begin{eqnarray}\label{eq:IRBCs}
 \bar{L}_{Ri}|_{\pi
 R_i^{\,-}}&=&-(Y^5_iH_i\bar{N}_{Ri}+\widetilde{Y}^5_i\widetilde{H}_i\bar{E}_{Ri})|_{\pi
 R_i^{\,-}},\\
 E_{Li}|_{\pi R_i^{\,-}}&=&\widetilde{Y}^5_i\widetilde{H}_iL_{Li}|_{\pi R_i^{\,-}},\\
 N_{Li}|_{\pi R_i^{\,-}}&=&Y^5_iH_iL_{Li}|_{\pi R_i^{\,-}},
\end{eqnarray}
\end{subequations}
where $\pi R_i^{\,-}\equiv \pi R_i-\epsilon$ for $\epsilon\rightarrow 0$
$(\epsilon>0)$. Note that $L_{Li}, E_{Ri},$ and $N_{Ri}$, are
continuous over the whole interval, including both endpoints,
i.e., in particular,  $E_{Ri}|_{\pi R_i^{\,-}}=E_{Ri}|_{\pi R_i}$ and $N_{Ri}|_{\pi R_i^{\,-}}=N_{Ri}|_{\pi R_i}$. (In contrast to this, $L_{Ri},E_{Li},$ and $N_{Li}$, are discontinuous at $y_i=\pi R_i$.)

\begin{figure}[t]
\begin{center}
\includegraphics*[width=0.6\textwidth]{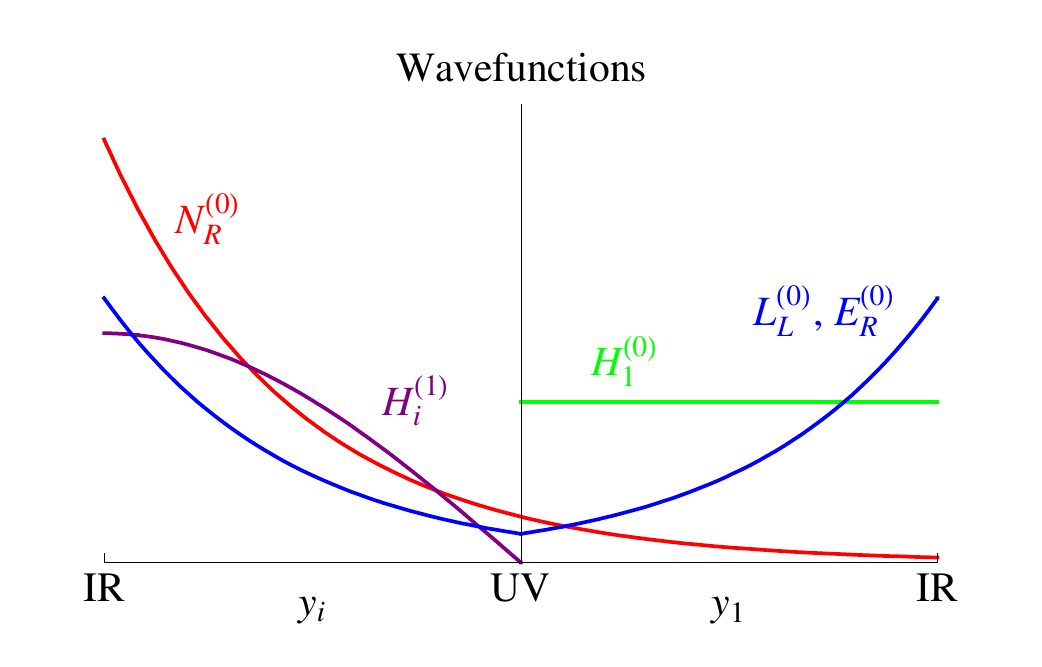}
\caption{Wavefunction profiles of the scalars and fermions in the
  throat geometry (schematic).  While the RH SM singlet neutrino zero modes $N_R^{(0)}$ are localized towards the IR branes of the throats 2 and 3 (combined here as
  $i=2,3$), all SM fermion
  zero modes $L_L^{(0)}$ and $E^{(0)}_R$ are symmetrically localized
  at all three IR branes of the throats. The Higgs zero mode lives
  only on the 1st throat giving a large overlap with all SM fermions
  but only a small overlap with the RH neutrino. The
  overlap of $N^{(0)}_R$ and the other fermions with the heavy scalars $H^{(1)}_{2,3}$, however, is large.}\label{fig:wavefunctions}
\end{center}
\end{figure}

In our model, we have the LH and RH fermion
zero modes $L_L^{(0)},E^{(0)}_R,$ and $N^{(0)}_R$, with wavefunctions
\begin{eqnarray}
L^{(0)}_L(x,y_i)&=&A_0^L\,\text{exp}(-m^L_iy_i)\,\ell_L(x),\nonumber\\
E^{(0)}_R(x,y_i)&=&B_0^E\,\text{exp}(m^E_iy_i)\,e_R(x),\\
N^{(0)}_R(x,y_i)&=&B_0^N\,\text{exp}(m^N_iy_i)\,\nu_R(x),\nonumber
\end{eqnarray}
that correspond, in the notation of Sec.~\ref{sec:Diracleptogenesis}, (up to a normalization) to $\ell_L$ and $e_R$ of the SM and to $\nu_R$. Here, $B^\Psi_0$ denotes the normalization factor for the RH fields, which is similar to $A^\Psi_0$ with only the signs in front of the bulk masses switched. We
choose for the fermions the bulk masses as given in Tab.~\ref{tab:bulkmasses}.
\begin{table}[t]
\begin{center}
\begin{tabular}{|c||c|c|c|}
\hline
&&&\vspace*{-4mm}\\
$\Psi_i$&$\text{sgn}(m^\Psi_1)$&$\text{sgn}(m^\Psi_2)$&$\text{sgn}(m^\Psi_3)$\\
\hline
$L_i$&$-1$&$-1$&$-1$\\
\hline
$E_i$&$+1$&$+1$&$+1$\\
\hline
$N_i$&$-1$&$+1$&$+1$\\
\hline
\end{tabular}
\caption{Signs of the bulk masses for the fields $L_i,E_i,$
  and $N_i$, on the different throats.
}\label{tab:bulkmasses}
\end{center}
\end{table}
As a consequence, the zero modes of the LH
lepton doublets $L^{(0)}_L$ and the RH charged leptons
$E^{(0)}_R$ become symmetrically localized at the three IR branes of the throats. In contrast to this, the zero modes of the RH
neutrinos $N^{(0)}_R$ are only localized towards the IR branes of the
throats 2 and 3 but are repelled from the IR brane of the 1st throat. Schematically, the wavefunctions of the fields
in the bulk, including the wavefunctions of the scalars, are depicted in Fig.~\ref{fig:wavefunctions}.
Note that $L^{(0)}_L$ and $E^{(0)}_R$ have a large overlap with the zero mode $H^{(0)}_1$ which generates the lepton masses after acquiring a nonzero VEV. The overlap of the
RH neutrinos $N^{(0)}_R$ with $H^{(0)}_1$ is, however, exponentially small, thereby suppressing the Dirac neutrino masses of the active neutrinos. The overlap of $N^{(0)}_R$ and $L^{(0)}_L$
with the higher KK-excitations $H^{(n)}_2$ and $H^{(n)}_3$, on the
other hand, is larger, giving larger Dirac neutrino Yukawa couplings
to the heavy scalars. We suppose for each particle species $\Psi$ that the bulk masses $m_i^\Psi$ are flavor diagonal and degenerate for a fixed throat number $i$. This could, e.g., be ensured by an $SU(3)$ flavor symmetry that is preserved on each throat but broken at the UV and IR branes. We will, however, not discuss further the bulk flavor symmetry and its breaking but assume from now on simply the flavor-diagonal structure of the bulk masses and their degeneracy on each throat.

In what follows, we will, for simplicity, go to the limit of strongly
localized fermion zero modes, i.e.~$m_i^\Psi\pi R_i\gg 1$. In the low-energy
effective theory, the lepton Yukawa couplings of the fermion zero modes
are then described by the Lagrangian
\begin{eqnarray}\label{eq:LYeff}
 \mathcal{L}^{Y}_\text{eff}&=&\sum_{a,b=1}^3\,\Big\{(Y_1)_{ab}H_1^{(0)}\bar{\ell}_{L,a}\nu_{R,b}+(\widetilde{Y}_1)_{ab}\widetilde{H}^{(0)}_1\bar{\ell}_{L,a}e_{R,b}\nonumber\\
&+&\sum_{i=2,3}\big[(Y_i)_{ab}H_i^{(1)}\bar{\ell}_{L,a}\nu_{R,b}+(\widetilde{Y}_i)_{ab}\widetilde{H}^{(1)}_i\bar{\ell}_{L,a}e_{R,b}
\big]\Big\}+\text{h.c.}+\dots,
\end{eqnarray}
where $Y_i\propto Y_i^5$ and $\widetilde{Y}_i\propto\widetilde{Y}^5_i$ are the Yukawa coupling matrices of the 4D theory obtained after integrating out the extra dimension. In (\ref{eq:LYeff}), we have included only the lightest KK scalars and restored the generation indices.  The 4D Yukawa coupling matrices satisfy relations similar to those in (\ref{eq:Yukawarelations}):
\begin{equation}\label{eq:4DYukawarelations}
Y_1\propto Y_2=P^L Y_3,\quad \widetilde{Y}_1=\widetilde{Y}_2= P^L \widetilde{Y}_3.
\end{equation}
Since the symmetry $D_1$ remains unbroken at the IR branes, the Yukawa coupling matrices $Y_1$ and $Y_{2,3}$ are, to leading
order, in the 4D theory related by an overall rescaling factor
\begin{equation}\label{eq:rescaling}
 Y_1=\frac{1}{\sqrt{2}}e^{-2\pi R\,|m^N|}\,Y_2=\frac{1}{\sqrt{2}}e^{-2\pi R\,|m^N|}\,P^L Y_3,
\end{equation}
where $R=R_1=R_2=R_3$ and $|m^N|=|m_1^N|=|m_2^N|=|m_3^N|$. This connects directly the
low-energy Yukawa coupling matrices $Y_1$, accessible to neutrino
oscillation experiments, with the Yukawa coupling matrices $Y_{2,3}$ that describe the interactions of the SM leptons with the heavy scalars $H^{(n)}_{2,3}$. Calling the rescaling factor  $F=\frac{1}{\sqrt{2}}\,\text{exp}\,(-2\pi R\,|m^N|)$, we will later choose $F\simeq 10^{-5}$ to obtain realistic Yukawa couplings that give the right size $\sim 10^{-1}\,\text{eV}$ for the observed neutrino masses, while enabling, at the same time, successful leptogenesis.

In $\mathcal{S}_\text{UV}$, we will assume the matrices
$\mathcal{K}^\Psi_{ij}$ to be
\begin{equation}\label{eq:Kij}
\mathcal{K}_{ij}^{\Psi}\propto\left(
\begin{matrix}
 \delta & -\delta\\
-1 & -1\\
1-\delta & 1+\delta
\end{matrix}
\right),
\end{equation}
where $\delta\ll 1$ is a small symmetry breaking parameter. From (\ref{eq:fermionBCs}), we thus have
at the UV brane $\bar{\Psi}_{L1}|_{y_1=0}=\bar{\Psi}_{L2}|_{y_2=0}=\bar{\Psi}_{L3}|_{y_3=0}$
(for $\Psi_{Li}$ as an example). We see that each field
$\Psi_{Li}$ makes up $1/3$ of the zero mode wavefunction. The matrices in (\ref{eq:Kij}) break
the symmetry $D_1$ but $D_2$ is only slightly broken.

At a finite temperature $T$, the heavy scalars $H_i^{(n)}$ receive additive thermal corrections to their mass-squares, which are, neglecting Yukawa interactions, to leading order given by \cite{Sher:1988mj}
\begin{equation}
\Delta M_{nT}^2(H_i)=\frac{3\lambda_i^{(n)}}{24}T^2,
\end{equation}
where $\lambda^{(n)}_i$ is the quartic self-coupling of the $n$th KK Higgs
field on the $i$th throat. The symmetry $D_2$ establishes
$\lambda_2^{(n)}=\lambda_3^{(n)}$, such that the corresponding leading
order thermal corrections to the mass-squared splittings $M_n^2(H_2)$
and $M_n^2(H_3)$ are zero. The fermion mass matrix
$\mathcal{K}_{ij}^\Psi$ in (\ref{eq:Kij}) breaks $D_1$ and $D_2$. But
since the Higgs fields $H_i$ do not couple to the symmetry breaking
terms at the UV brane, $\mathcal{S}_\text{UV}$ will only produce an unobservable shift in the potential without changing the thermal
masses of the scalars \cite{Sher:1988mj}.

\section{Leptogenesis}\label{sec:Leptonasymmetry}
\subsection{Bounds on Yukawa Couplings}
\begin{figure}
\begin{center}
\includegraphics*[width=0.50\textwidth]{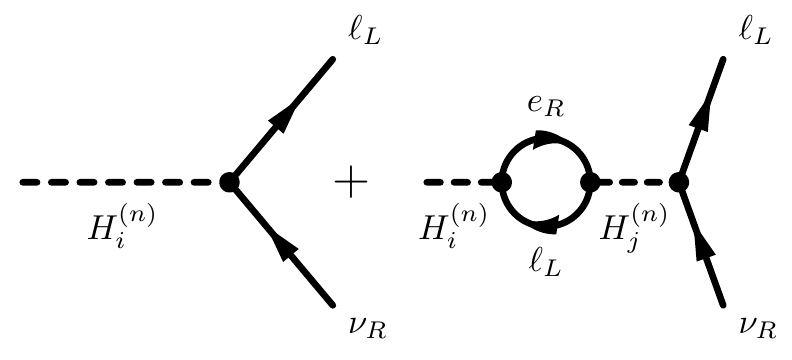}
\caption{Decay amplitudes generating the decay asymmetries
  $\epsilon_i^{(n)}$ ($i=2,3$).}\label{fig:decay}
\end{center}
\end{figure}

Let us now see how in our model the observed baryon asymmetry is
generated via Dirac leptogenesis
\cite{Dick:1999je} through
the decay of the heavy scalar doublets $H_i^{(n)}$ ($i=2,3$). The scalars decay via
\begin{equation}\label{eq:decays}
 H_{2,3}^{(n)}\rightarrow \ell_L+\bar \nu_R,\bar \ell_L+e_R,
\end{equation}
involving the Yukawa coupling matrices $Y_{2,3}$ (see Fig.~\ref{fig:decay}).
The allowed parameter space of the KK masses $M_n(H_{2,3})$ and the corresponding neutrino Yukawa coupling matrices $Y_{2,3}$ is restricted by several bounds. First of all, according to Sakharov's third condition, the asymmetry generating processes have
to be out of equilibrium, i.e.~in our scenario, the KK scalars have to
decay at temperatures $T\lesssim M_1(H_i)$. Dirac
leptogenesis, on the other hand, requires that all relevant decay
processes have to take place at energies above the scale of
EWSB $T_c$, i.e.~as long as the sphaleronic processes are in thermal equilibrium \cite{sphalerons}. We thus have a time limit  $\Delta t_\text{D}$ for the relevant decays of the lowest scalar KK excitations
\begin{eqnarray}
\Delta t_\text{D}=t(T_c)-t(M_1(H_i))
=0.30\times\frac{M_\text{Pl}}{\sqrt{g_\ast}}\,\bigg(\frac{1}{T_c^2}-\frac{1}{M^2_1(H_i)}\bigg),
\end{eqnarray}
where the time $t$ is given by $t=\frac{1}{2}\,H^{-1}$ with the Hubble parameter $H=1.66\times\sqrt{g_\ast}\frac{T^2}{M_\text{Pl}}$ and the factor 
$g_*$ denoting the number of accessible relativistic degrees of freedom at temperature $T$.

The time $\Delta t_\text{D}$ has to be compared with the life time $\tau_\nu$ for
decays into RH neutrinos:
\begin{equation}
\tau_{\nu}\equiv \Gamma_{\nu}^{-1}
=\frac{16 \pi}{\text{tr}(Y_iY_i^\dagger)\, M_1(H_i)}.
\end{equation}
Since the decays take place at temperatures that are small compared to the scalar
masses, the Lorentz gamma factor can be neglected. For the asymmetry generating processes to take place we obviously need $\Delta t_\text{D}\gtrsim\tau_\nu$.

Moreover, for temperatures $T>T_c$, in order to avoid LR equilibration by scattering processes, it is necessary that $H(T)\gtrsim \Gamma_\text{S}(T)$, where $\Gamma_\text{S}(T)=n_\text{eq}\langle\sigma_S|v|\rangle$ denotes the rate of $\nu_R$ annihilating scattering processes, $n_\text{eq}$ the equilibrium number density of target particles, and $\langle\sigma_S|v|\rangle$ is the thermally averaged scattering cross section. For $H_1$, the dominant contribution from scattering processes at $T_c$ can be
approximated as $\Gamma_\text{S}(T_c)\sim \lambda^2(Y_1)^2\,T_c$, where $\lambda\sim1$ is a typical gauge or Yukawa coupling. Consequently, the Yukawa couplings roughly satisy
\begin{equation}
Y_1\lesssim\sqrt{\frac{T_c}{M_\text{Pl}}}\sim 10^{-8},
\end{equation}
where we have neglected the generation indices. Observe that this condition is fulfilled by Dirac neutrinos with Yukawa couplings of the order
$Y_1\sim 10^{-12}$ that give neutrino masses $\sim 10^{-1}\,\text{eV}$ consistent with observation. For the heavy scalars $H^{(1)}_{2,3}$, the dominant scattering process takes the form \cite{Kolb:1990vq}
\begin{equation}
\Gamma_\text{S}(T_c)\sim
 (\widetilde{Y}_iY_i)^2\,T_c^5/M_1^4(H_i).
\end{equation}
Setting the charged lepton Yukawa couplings $\widetilde{Y}_i\sim 1$, the bound on the $Y_i$ becomes
\begin{equation}
Y_i\lesssim\sqrt{\frac{M_1^4(H_i)}{T_c^3M_\text{Pl}}}.
\end{equation}
In order to measure the effectiveness of decays at $T\sim
M_1(H_i)$, we introduce the quantity
\begin{equation}
K=\frac{\Gamma(H_i)}{2H(M_1(H_i))}=\frac{(Y_i)^2}{1.66\times32\pi\sqrt{g_\ast}}\frac{M_\text{Pl}}{M_1(H_i)}.
\end{equation}
For $K\ll1$, we are in a regime of pure ``drift and decay'' \cite{Kolb:1990vq}, called the weak
wash-out regime, i.e.~inverse decays are strongly suppressed and cannot
erase the produced asymmetry.

\begin{figure}
\begin{center}
\includegraphics*[viewport = 0 0 360 200, width=0.7\textwidth]{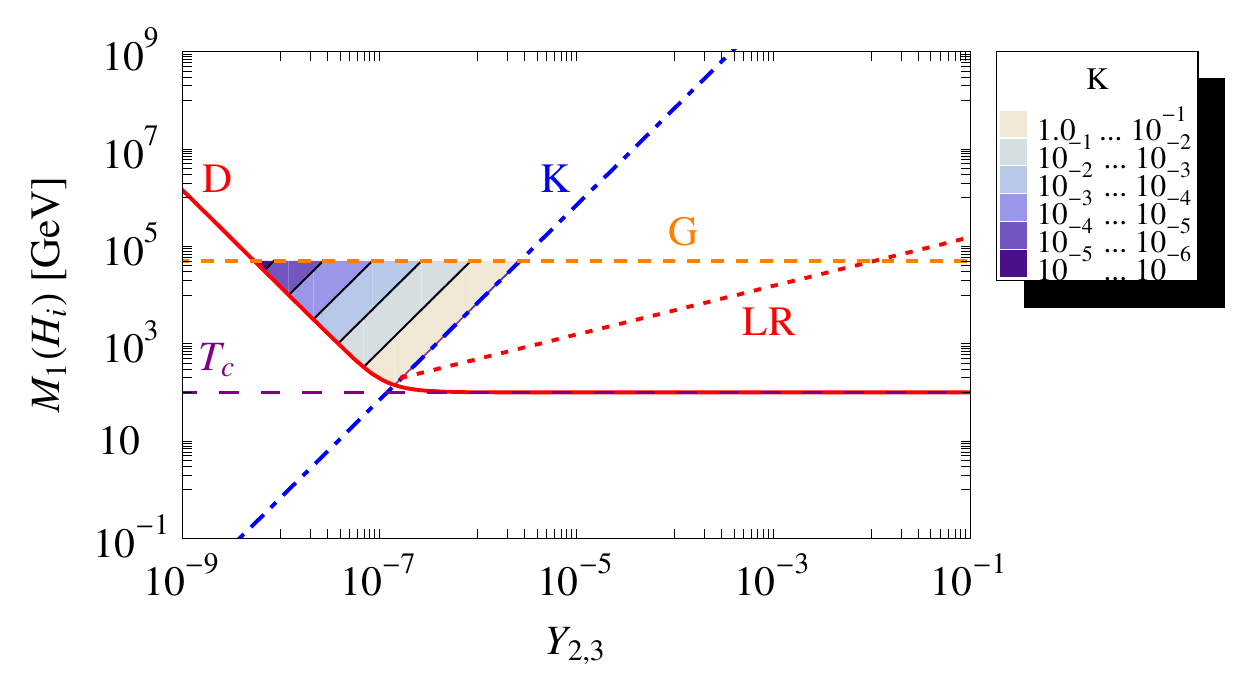}
\caption{Summary of bounds on the Dirac neutrino Yukawa couplings
  $Y_{2,3}$ to the heavy Higgs doublets $H_2$ and $H_3$. The
  lines labeled by D, K, and LR, denote the exclusion regions for
  late decays (D), the weak washout regime (K), and LR equilibration
  (LR). Above $T_c$, sphaleron processes
  are in thermal equilibrium and $T_\text{RH}$ is restricted below G by KK 
  graviton decays. The preferred
  region considered here is $Y_{2,3}\sim 10^{-7}$ and $M_1(H_{2,3})\lesssim 50\,\text{TeV}$.}\label{fig:bounds}
\end{center}
\end{figure}

An upper limit on the KK masses $M_n(H_i)$ is set by the graviton
bound \cite{Eisele:2007ws, graviton}, i.e.~by the requirement that late decays should not spoil the production of light elements during Big Bang nucleosynthesis (BBN) at $T_\text{BBN}\sim0.3\,\text{MeV}$. On dimensional grounds, the KK graviton (G) decay rates take in our model the form
\begin{equation}\label{eq:Grate}
\Gamma_{G^{(n)}}\sim n_\text{D}\,\frac{M^3_n(G)}{M_\text{Pl}^2},
\end{equation}
where $n_\text{D}$ denotes the number of decay channels and $M_n(G)$ is the mass of the $n$th graviton excitation.
Note that due to the wavefunction profiles in our model the KK number is not conserved. From (\ref{eq:Grate}) we see that the KK graviton life time strongly depends on the KK masses, i.e.~the compactification scale. Thus, we find two different parameter ranges: For $M_1(G)\gtrsim50\,\text{TeV}$, all KK excitations of the gravitons decay before the era of BBN and thus no further restrictions arise \cite{Eisele:2007ws}. However, for the energy range we are particularly interested in, i.e.~$M_1(G)\lesssim50\,\text{TeV}$, KK graviton decays lead to a bound on the reheating temperature $T_\text{RH}$ \cite{graviton}. In this case, $T_\text{RH}$ must be close to the compactification scale and, consequently, only a few KK modes can come on-shell. In Fig.~\ref{fig:bounds}, we have summarized the bounds on
the Dirac neutrino Yukawa couplings to the heavy Higgs doublets. The
lines labeled by D, K, and LR, denote the exclusion regions for late decays (D), weak washout regime (K), and LR equilibration
(LR). The allowed parameter space lies above the lines $T_c$, D, K, and LR. The preferred region is therefore around $Y_{2,3}\simeq
10^{-7}$ and $M_1(H_{2,3})\simeq 1\,\text{TeV}\dots 50\,\text{TeV}$.\footnote{For $Y_2\sim 10^{-12}$
  of the order the Dirac Yukawa couplings of the active neutrinos, we
  would need to go to a resonant limit with very tiny relative scalar
  mass-squared splittings $\sim 10^{-20}$.} The neutrino Yukawa couplings to the heavy Higgs fields are therefore
small but still by a factor $\sim 10^5$ larger than the neutrino
Yukawa couplings $Y_1\sim 10^{-12}$ to the SM Higgs $H^{(0)}_1$, generating the observed neutrino masses. In the flat limit, the mass
range for $M_1(H_{2,3})$ translates into the range $R^{-1}\simeq
1...50\,\text{TeV}$ for the compactification scale, implying a fundamental Planck scale of the order
$M_*=(M_\text{Pl}^2/R)^\frac{1}{3}\simeq 10^{13}\dots
10^{14}\,\text{GeV}$.

\subsection{Lepton Asymmetry}
The lepton asymmetry is generated by the interference between the tree-level and
one-loop amplitudes shown in Fig.~\ref{fig:decay}. The one-loop amplitude must involve on-shell LH lepton doublets
and RH charged leptons. The decay of the KK mode $H_2^{(n)}$, e.g., leads to the decay asymmetry \cite{Nanopoulos:1979gx,Liu:1993tg}
\begin{eqnarray}\label{eq:epsilon2}
\epsilon^{(n)}_2&=&\frac{\Gamma(H_2^{(n)}\rightarrow \ell_l\bar\nu_R )-\Gamma(\bar H_2^{(n)}\rightarrow \bar\ell_l\nu_R)}{\Gamma(H_2^{(n)})+\Gamma(\bar H_2^{(n)})}
\nonumber\\
&\approx&
\frac{
\text{Im}\,[\text{tr}(Y_2^\dagger Y_3)\,\text{tr}(\widetilde{Y}_2^\dagger\widetilde{Y}_3)]}{8\pi\,\text{tr}(\widetilde{Y}_2^\dagger \widetilde{Y}_2)}\,\frac{M^2_n(H_2)}{M^2_n(H_3)-M^2_n(H_2)},
\end{eqnarray}
where we have used the fact that the dominant contribution comes from the pair of
KK states with the same level number $n$ \cite{Pilaftsis:1999jk}. In
(\ref{eq:epsilon2}), we
have only taken the self-energy contributions into account, since additional vertex
corrections can be neglected in the resonant limit. Note
that our expression for the resonantly enhanced lepton asymmetry in
(\ref{eq:epsilon2}) corrects the result in \cite{Dick:1999je} in two ways: We have (i) included a complex conjugation of the Yukawa coupling matrix $Y_2$ and have (ii) taken in
the numerator the product of two traces instead of a single trace.

Equation (\ref{eq:epsilon2}) holds as long as $M_n(H_3)-M_n(H_2)\gg\Gamma_n(H_2)$, which is satisfied for our range of parameters with small Yukawa couplings. Similarly, the decay of $H_3^{(n)}$ leads to a decay asymmetry
$\epsilon^{(n)}_3$ which is obtained from the expression for
$\epsilon^{(n)}_2$ by interchanging in (\ref{eq:epsilon2}) the Yukawa
coupling matrices $Y_2\leftrightarrow Y_3$ and the Higgs fields
$H_2\leftrightarrow H_3$. For $(Y_2)_{ab},(Y_3)_{ab}\lesssim 10^{-7}$, a scalar mass-squared
splitting of the order $\sim 10^{-8}$ as given in (\ref{eq:splitting})
produces a total decay asymmetry
$\epsilon_\text{total}^{(n)}=\epsilon_2^{(n)}+\epsilon_3^{(n)}$
of the order $\epsilon_\text{total}^{(n)}\sim -10^{-8}$. For this
choice of parameters, an equal number $n_{H_{i}^{(1)}}$ of $H^{(1)}_i$ and $\bar{H}^{(1)}_i$ ($i=2,3$) leads to a net number density in the RH neutrino sector
$n_{\nu_R}^{(n)}\simeq
4\,\epsilon^{(n)}_\text{total}n_{H_{i}^{(n)}}$. The out-of equilibrium
decays in the ``drift and decay'' limit \cite{Kolb:1990vq}, i.e.~$n_{H_i^{(n)}}\sim n_\gamma$, thus yield a neutrino number to entropy ratio
\begin{equation}
Y_\nu=
\sum_{n=1}^{n_\text{max}}\frac{n_{\nu_R}^{(n)}}{s^{(n)}}\sim\sum_{n=1}^{n_\text{max}}\frac{4\,\epsilon_\text{total}^{(n)}}{g_\ast^{(n)}}.\label{eq:Ynu}
\end{equation}
In (\ref{eq:Ynu}), the dominant contribution to the asymmetry is generated around the
energy at which $H_{2,3}^{(n)}$ drop out of thermal equilibrium such
that the entropy is given by $s^{(n)}=g_\ast^{(n)} n_\gamma$, where
$g_*^{(n)}$ is the number of relativistic degrees of freedom at
$T\sim M_n(H_{2,3})$. In the standard model, we have
$g_\ast\sim100$.  This is altered in the 5D model by taking the
additional KK excitations into account, giving
$g_\ast^{(n)}=g_\ast^{(0)}+g_\ast^\text{KK}\Theta(T-R^{-1})\,TR$, where
$\Theta(x)=1$, for $x\geq0$, and $\Theta(x)=0$, for $x<0$. The number of relativistic degrees of freedom for the zero modes is
$g_\ast^{(0)}\sim100$ and we take also $g_\ast^\text{KK}\sim g_\ast^{(0)}$.
The asymmetry $\epsilon_\text{total}^{(n)}$ is independent of $n$,
since the scalar mass-squared splittings as well as the expression for the Yukawa
couplings are independent from $n$. Thus, we can approximate\footnote{For a large number $n_\text{max}$ of scalar KK excitations the sum over $1/g^{(n)}_\ast$ can be approximated logarithmically \cite{Pilaftsis:1999jk} by $\sum_{n=1}^{n_\text{max}}(g_\ast^{(0)}+n\, g_\ast^\text{KK})^{-1}\approx\ln\,\Big(\frac{g_\ast+n_\text{max}g_\ast^\text{KK}}{g_\ast+g_\ast^\text{KK}}\Big)$ and we thus obtain a factor $\mathcal{O}(1)$ for a large range of energies instead of a factor $\sim 10^{-2}$, as in the case here of only a few KK states.}
the final asymmetry by
\begin{equation}
Y_\nu\sim\frac{\epsilon_\text{total}^{(1)}}{g_\ast}\sim -10^{-10}.
\end{equation}
An analysis of chemical potentials \cite{Harvey:1990qw} reveals that
for initial $B-L=0$ and all the heavy KK excitations decaying out of equilibrium,
the number densities of baryons and leptons are
related to the number density of RH neutrinos by
$n_B=n_L=-\frac{28}{79}\,n_{\nu_R}$ \cite{Dick:1999je}. This means
that $Y_\nu$ is converted by sphaleron
processes into a baryon asymmetry $Y_B\sim -Y_\nu$ of the
order the observed value $Y_B=(8.62\pm0.27)\times10^{-11}$ \cite{Amsler:2008zzb}.

\subsection{Correlation of Low-Energy Parameters}\label{sec:correlations}
The Yukawa coupling matrices appearing in the decay asymmetries
$\epsilon_2^{(n)}$ and $\epsilon^{(n)}_3$ exhibit the important
feature that they are related to the Yukawa couplings of the low-energy theory by the exchange symmetries $D_1$ and $D_2$ in (\ref{eq:D1}) and (\ref{eq:D2}). The BAU in our model becomes therefore connected with the low-energy neutrino masses, mixing angles, and the Dirac
CP phase observable in neutrino oscillations. In the basis where the charged lepton mass matrix is diagonal, the trace over the neutrino Yukawa couplings in (\ref{eq:epsilon2}) takes the form $\text{tr}(Y_2^\dagger Y_3)=\text{tr}[U_{\text{PMNS}}Y_3^{\text{diag}} Y_3^{\text{diag}\,\dagger}U^\dagger_{\text{PMNS}}P^{L\,^\dagger}]$. Inserting the Yukawa couplings in (\ref{eq:4DYukawarelations})
into the expression for $\epsilon^{(n)}_{2,3}$, one can then study the BAU as a function of the solar, atmospheric, and reactor mixing
angles $\theta_{12},\theta_{23},$ and $\theta_{13}$, and the Dirac CP phase $\delta$, of the low-energy leptonic mixing matrix $U_{\text{PMNS}}$ \cite{PMNS}.\footnote{For a discussion of possible forms of Dirac neutrino mass matrices see also \cite{Hagedorn:2005kz}.}
\begin{figure}[!ht]
\begin{center}
\subfloat[\hspace{-1.9cm}(a)]{
\includegraphics*[viewport = 0 0 360 245, width=0.487\textwidth]{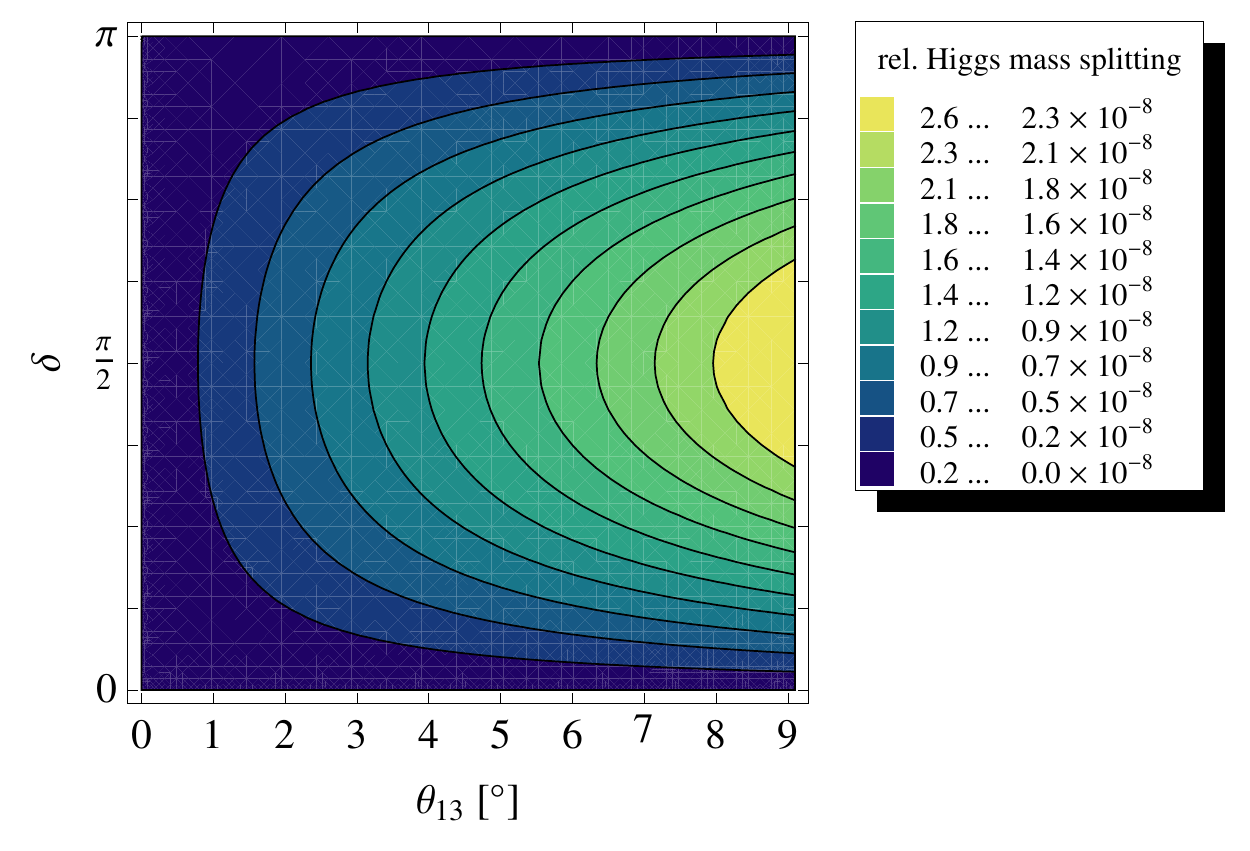}}
\subfloat[\hspace{-1.9cm}(b)]{
\includegraphics*[viewport = 0 0 360 245, width=0.487\textwidth]{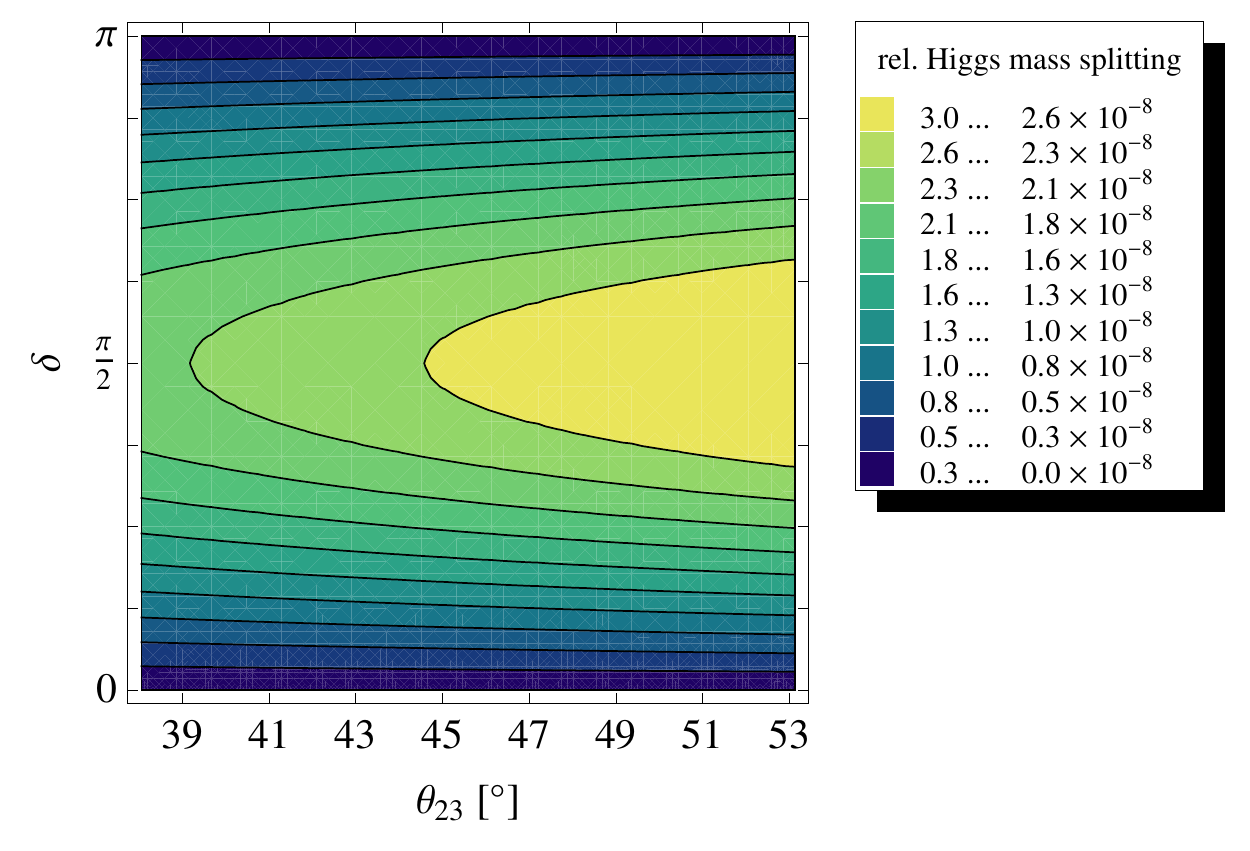}}\\
\subfloat[\hspace{-1.9cm}(c)]{
\includegraphics*[viewport = 0 0 360 245, width=0.487\textwidth]{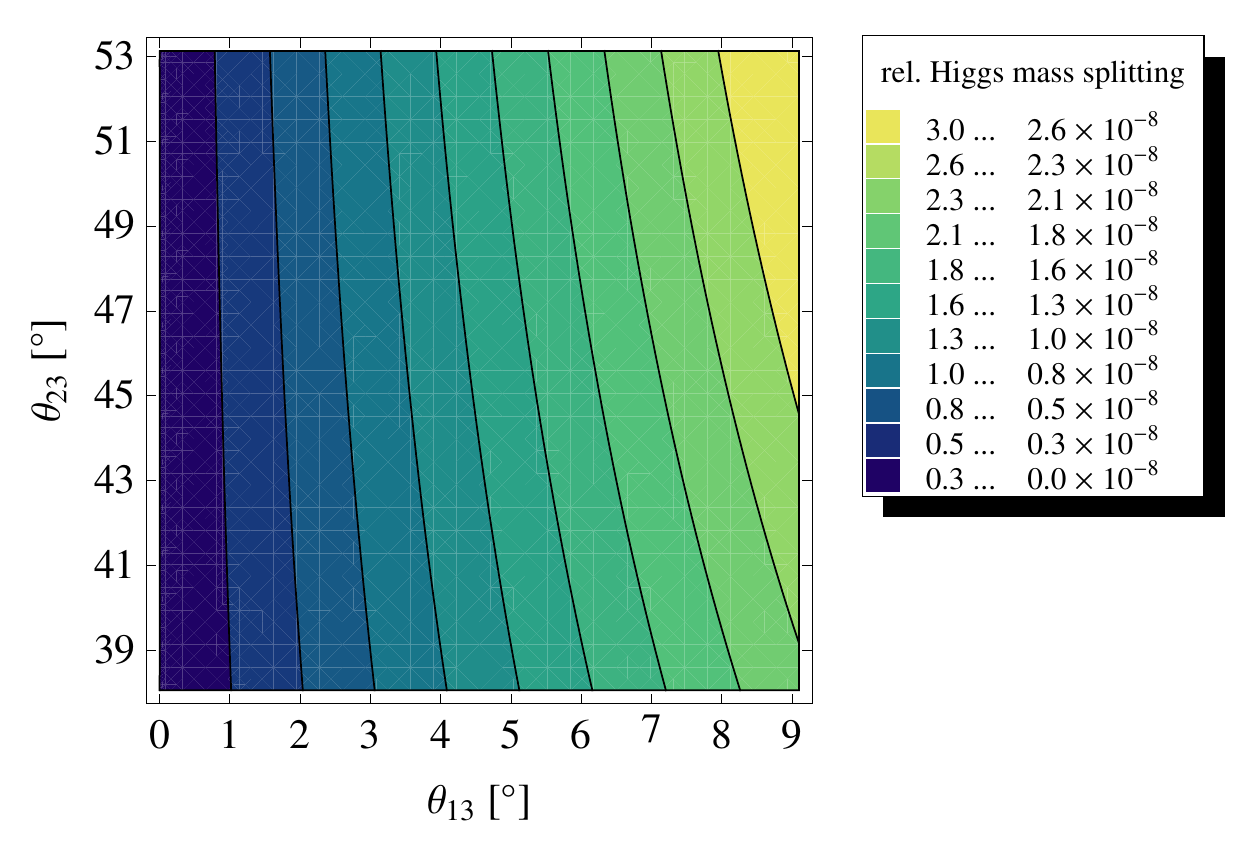}}
\caption{Correlations between $\theta_{13}$ and $\delta$ (a), $\theta_{23}$ and $\delta$ (b), and $\theta_{13}$ and $\theta_{23}$ (c), for $P^L$ in (\ref{eq:Pij}) as a function of the relative Higgs mass(-squared) splitting. In (a), (b), and (c), we have assumed the values $\theta_{23}=45.0^\circ$, $\theta_{13}=9.1^\circ$, and $\delta=\pi/2$, respectively. }\label{fig:D24}
\end{center}
\end{figure}\\

In Fig.~\ref{fig:D24}, we show the correlations between the reactor angle $\theta_{13}$, the atmospheric angle $\theta_{23}$, and the Dirac CP phase $\delta$, for the matrix representation $P^L$ in (\ref{eq:Pij}) that generates a $Z_4$ subgroup of $\Delta(24)$. The correlations are given as a function of the Higgs mass(-squared) splitting in (\ref{eq:splitting}) for the lowest KK excitation ($n=1$). The solar angle and BAU are fixed at their best-fit values $\theta_{12}=33.2^\circ$ and $Y_B=8.62\times10^{-11}$. We have assumed a normal hierarchical neutrino mass spectrum with mass ratios $m_1:m_2:m_3=0.04:0.2:1$ (similar results can, however, also be obtained for an inverted neutrino mass hierarchy). Denoting by $Y_i^\text{diag}$ the matrix obtained after diagonalization of the 4D Dirac neutrino Yukawa coupling matrix $Y_i$, we have, in Fig.~\ref{fig:D24}, taken  $(Y_2^\text{diag})_{33}=(Y_3 ^\text{diag})_{33}=10^{-7}$, while the mass splitting of the decaying scalars has been varied in the range $0...10^{-8}$. The region for $\delta>\pi$ is unphysical, since it would lead to a wrong sing of the BAU. In the appendix, we present the results for the same parameters with $P^L$ taken as a matrix representation of elements of the groups $\Delta(27)$ and $\Delta(54)$. The correlations between the low-energy lepton mass and mixing parameters make our model testable at future neutrino oscillation experiments such as
Double Chooz \cite{Huber:2006vr}, T2HK \cite{Huber:2005jk}, or a neutrino factory \cite{Huber:2003ak}.

\section{Summary and Conclusions}\label{sec:Summary}
In this paper, we have presented a model for resonant Dirac
leptogenesis on a 5D flat multi-throat background. The baryon
asymmetry is generated by the decay of heavy scalars, which are copies of the SM Higgs. The throats which are subject to discrete exchange symmetries allow to solve several possible shortcomings of the original scenario for Dirac leptogenesis. First, the model provides an origin of the heavy decaying scalars as KK excitations of 5D Higgs fields. Second, the exchange symmetries protect a near mass degeneracy of the scalars which leads to resonant decays. This enables Dirac leptogenesis at energy scales as low as $1\,\text{TeV}\dots 50\,\text{TeV}$ that may be in reach of a collider. Third, the discrete symmetries, which are broken in the bulk, connect the observed BAU with the Yukawa couplings of the low-energy theory. This leads in our model to non-trivial correlations between the lepton mixing parameters. We have studied the dependence of the BAU on the atmospheric angle, the reactor angle, and the Dirac CP phase for several discrete group representations and found strong correlations between the mixing angles and the CP phase. This makes our model testable at future neutrino oscillation experiments such as neutrino factories.

It would be interesting, e.g., to consider in more detail the Boltzmann equations for our model, to study the throat stabilization necessary to ensure the resonant decays, and to investigate possible collider implications.

\section*{Acknowledgements}
We would like to thank M.T.~Eisele, H.~Murayama, and R.~R\"uckl, for very useful discussions. A.B. is supported by Research Training Group 1147 ``{\it Theoretical
  Astrophysics and Particle Physics}''of Deutsche
  Forschungsgemeinschaft. G.S. was supported by the Federal Ministry of Education and Research (BMBF) under contract number 05HT6WWA.

\appendix

\section{Correlations for Elements of $\Delta(27)$ and $\Delta(54)$}
\begin{figure}[ht]
\begin{center}
\includegraphics*[viewport = 0 0 360 245, width=0.6\textwidth]{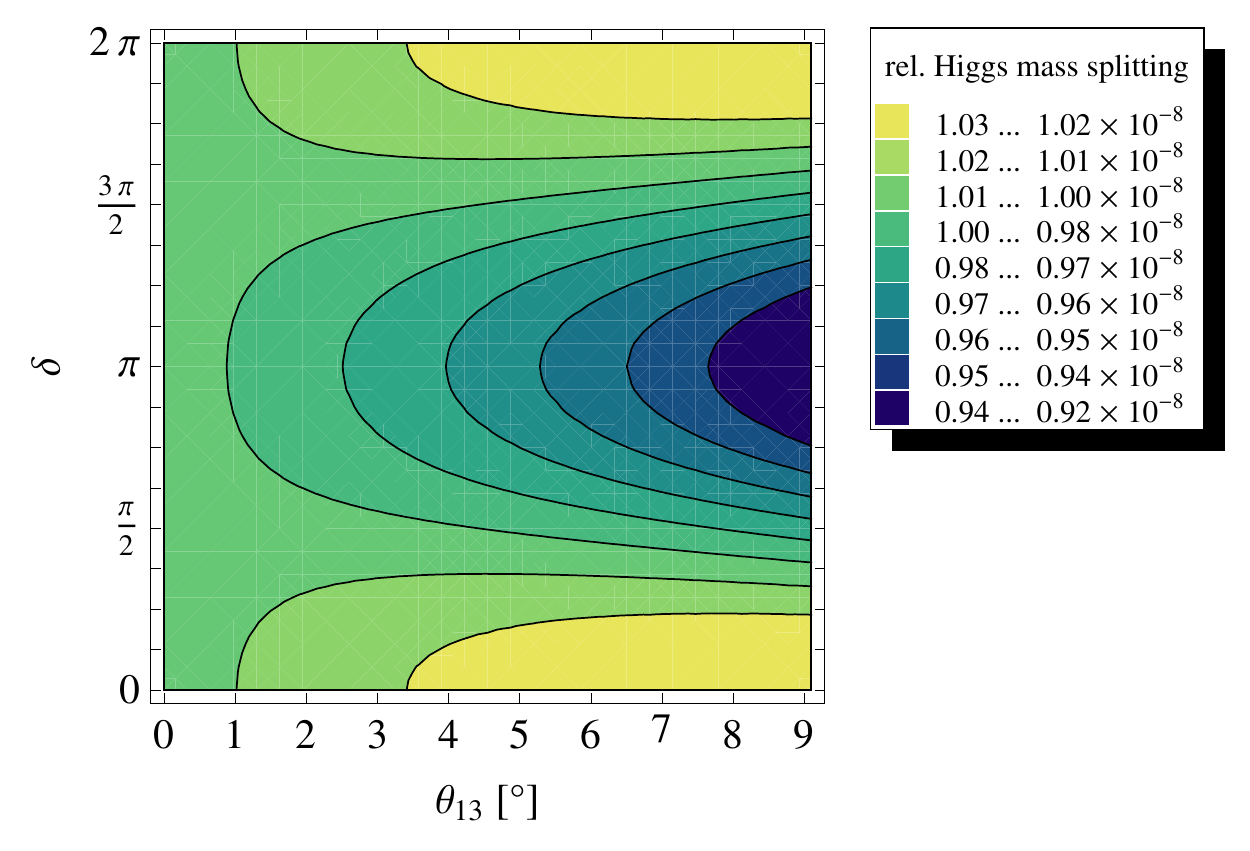}
\caption{Correlation between the reactor angle $\theta_{13}$ and the CP phase $\delta$, for $\theta_{23}=42.1^\circ$ and the matrix $P^L$ in (\ref{eq:PD27}).}\label{fig:D27}
\end{center}
\end{figure}

\begin{figure}[!ht]
\begin{center}
\subfloat[\hspace{-1.9cm}(a)]{\label{fig:D27th13da}
\includegraphics*[viewport = 0 0 360 245, width=0.487\textwidth]{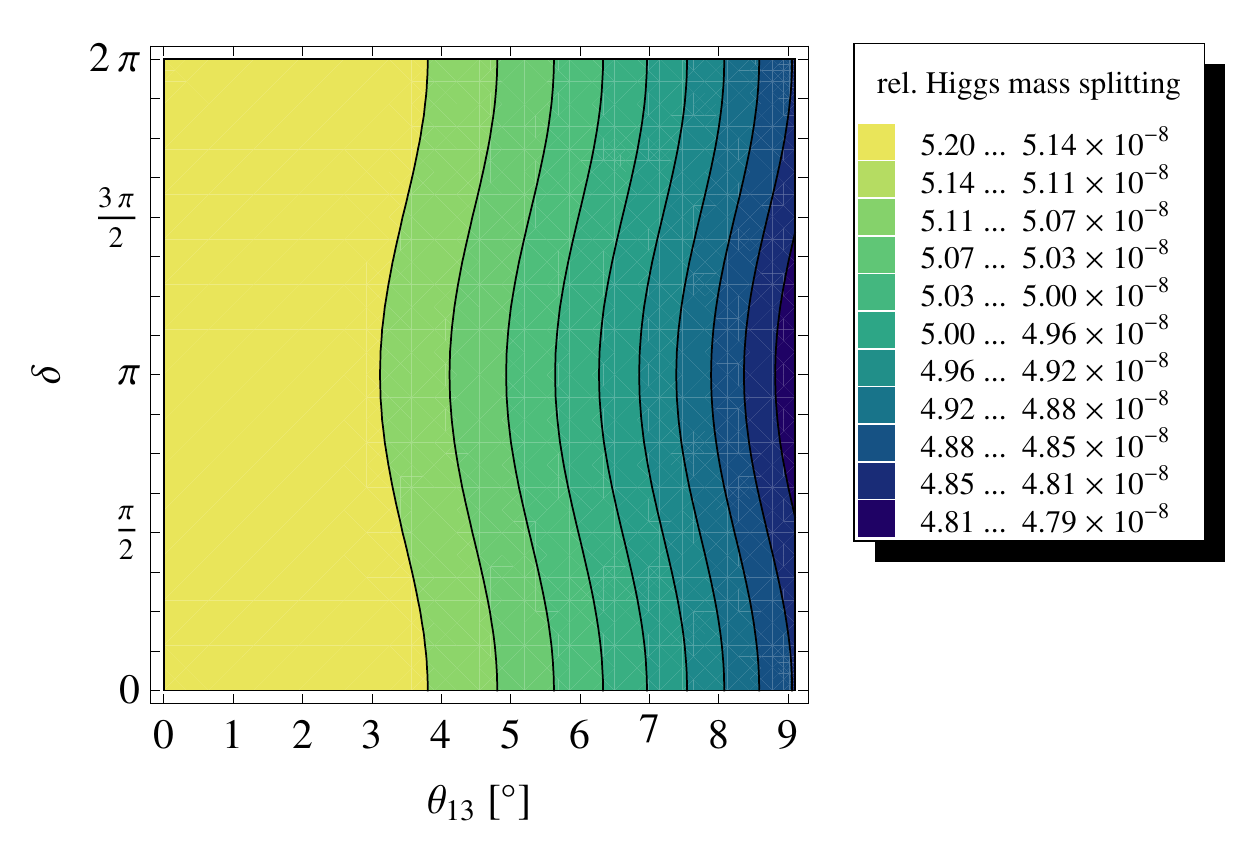}}
\subfloat[\hspace{-1.9cm}(b)]{\label{fig:D27th13th23a}
\includegraphics*[viewport = 0 0 360 245, width=0.487\textwidth]{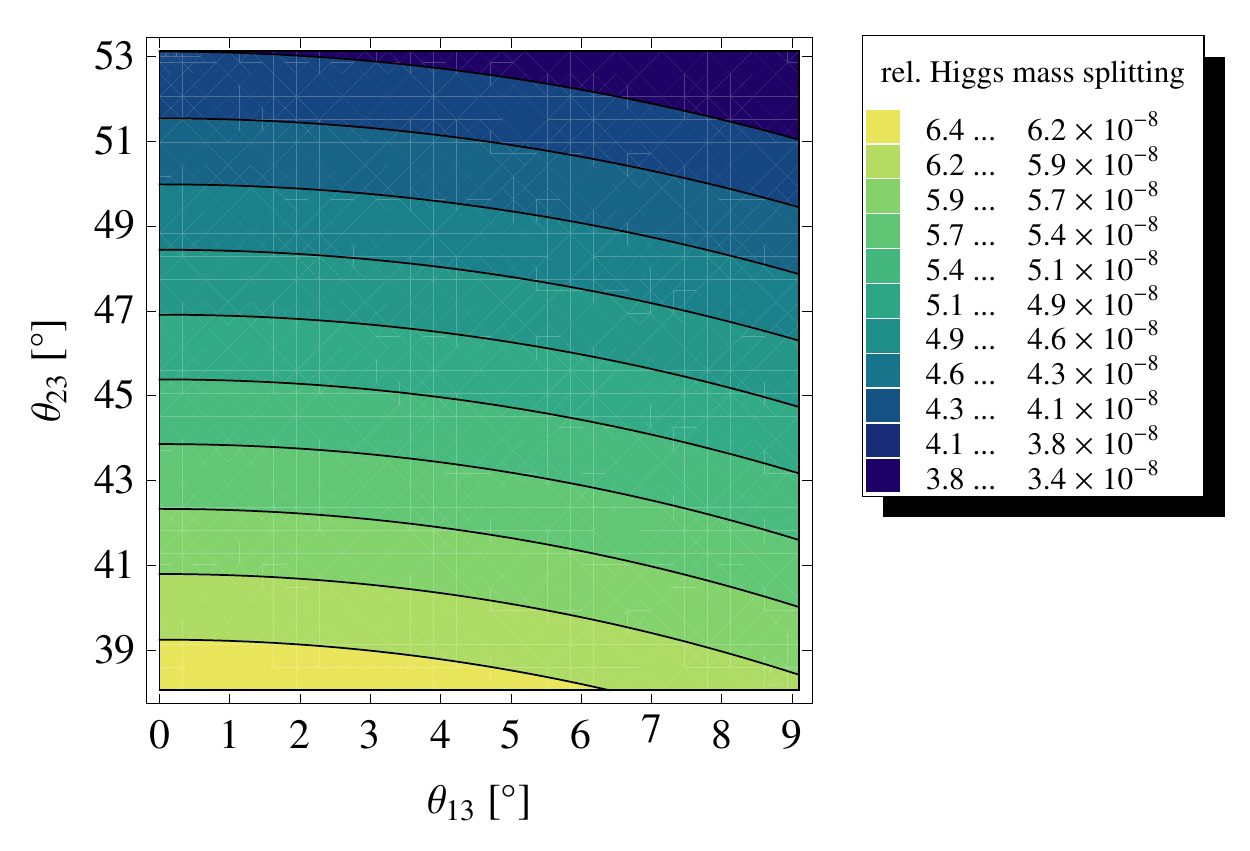}}
\caption{Correlations between $\theta_{13}$ and $\delta$ (a) and between $\theta_{13}$ and $\theta_{23}$ (b) for $P^L$ as in (\ref{eq:PD27a}). In (a) and (b) we have set $\theta_{23}=45.0^\circ$ and $\delta=\pi/2$, respectively. }\label{fig:D27a}
\end{center}
\end{figure}

Let us now present further results for the correlations between the low-energy lepton mixing parameters, when the matrix $P^L$ in (\ref{eq:Pij}) is a matrix representation of a generator of a cyclic subgroup of the groups $\Delta(27)$ or $\Delta(54)$ (for a discussion of $\Delta(54)$ as a flavor symmetry, see \cite{Ishimori:2008uc}). In what follows, we will, as in Fig.~\ref{fig:D24}, set throughout the solar angle and the BAU equal to their best fit values $\theta_{12}=33.2^\circ$ and $Y_B=8.62\times10^{-11}$. Moreover, the neutrino Yukawa couplings are chosen as in Sec.~\ref{sec:correlations}. As a first example, let us consider for $P^L$ the following matrix representation:
\begin{equation}\label{eq:PD27}
P^L=
\left(\begin{matrix}
\omega & 0 & 0\\
0 & 1 & 0\\
0 & 0 & \omega^2
\end{matrix}
\right),
\end{equation}
where $\omega=\exp(2\pi\text{i}/3)$. The matrix $P^L$ is taken from the class $3\,C_1^{(1,1)}$ \cite{Luhn:2007uq} or $C_9$ \cite{Ma:2006ip} of $\Delta(27)$ and generates a $Z_3$ symmetry. Fig.~\ref{fig:D27} shows the correlation between $\theta_{13}$ and $\delta$ for this choice of $P^L$.
The Higgs mass(-squared) splitting is defined as for Fig.~\ref{fig:D24}. Note in Fig.~\ref{fig:D27} that we have taken for the atmospheric angle the value $\theta_{23}=42.1^\circ$ (for $\theta_{23}=45^\circ$ the BAU would become further suppressed by two orders of magnitude due to an accidental cancelation). We show in Fig.~\ref{fig:D27} only the correlation between $\theta_{13}$ and $\delta$ since it is much stronger than the dependencies of these parameters on $\theta_{23}$. Radically different results are obtained for another group element from the same class of $\Delta(27)$ with matrix representation
\begin{equation}\label{eq:PD27a}
P^L=
\left(\begin{matrix}
\omega^2 & 0 & 0\\
0 & \omega & 0\\
0 & 0 & 1
\end{matrix}
\right).
\end{equation}
This matrix leads to the correlations between low-energy parameters shown in Fig.~\ref{fig:D27a}. We observe that in this case there is no strong dependence of $\theta_{13}$ on $\delta$.
\begin{figure}[!ht]
\begin{center}
\subfloat[\hspace{-1.9cm}(a)]{\label{fig:D54th13d}
\includegraphics*[viewport = 0 0 360 245, width=0.487\textwidth]{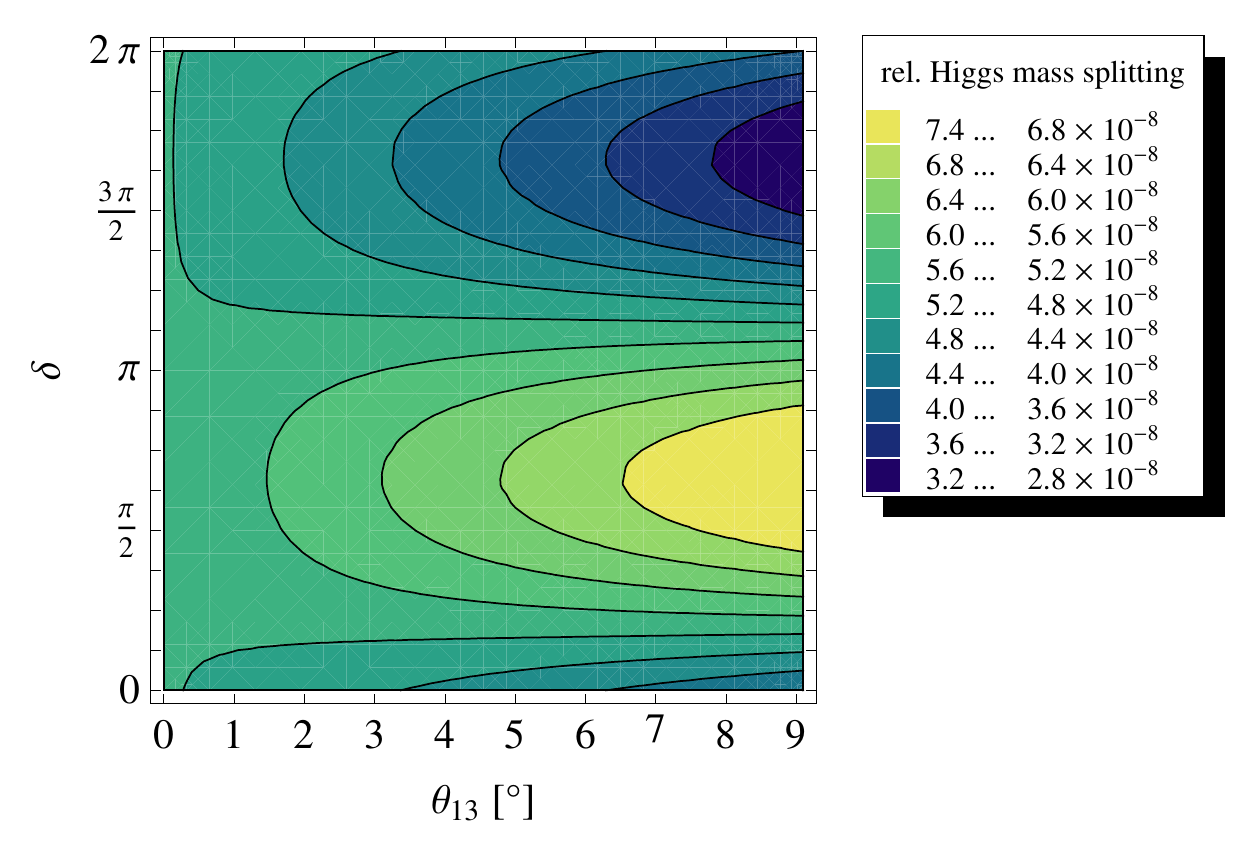}}
\subfloat[\hspace{-1.9cm}(b)]{\label{fig:D54th23d}
\includegraphics*[viewport = 0 0 360 245, width=0.487\textwidth]{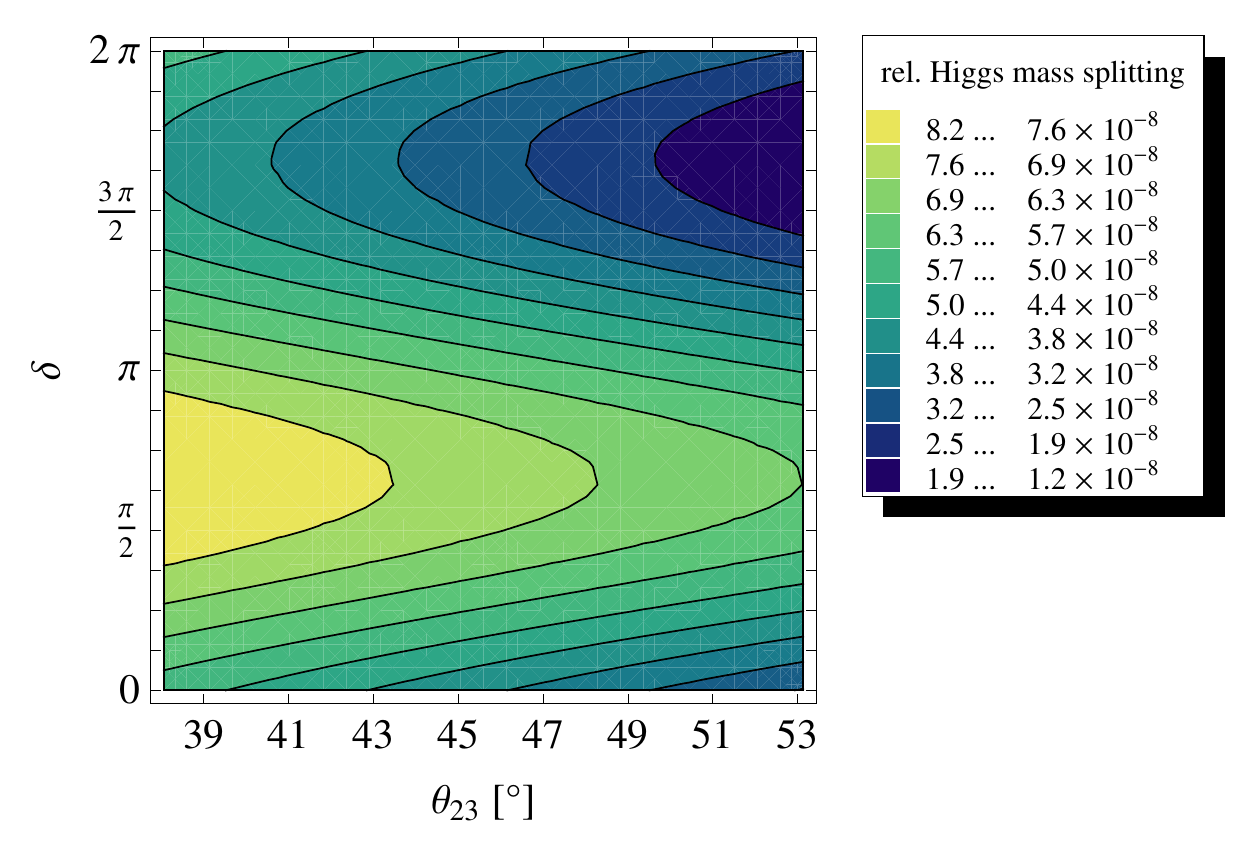}}\\
\subfloat[\hspace{-1.9cm}(c)]{\label{fig:D54th13th23}
\includegraphics*[viewport = 0 0 360 245, width=0.487\textwidth]{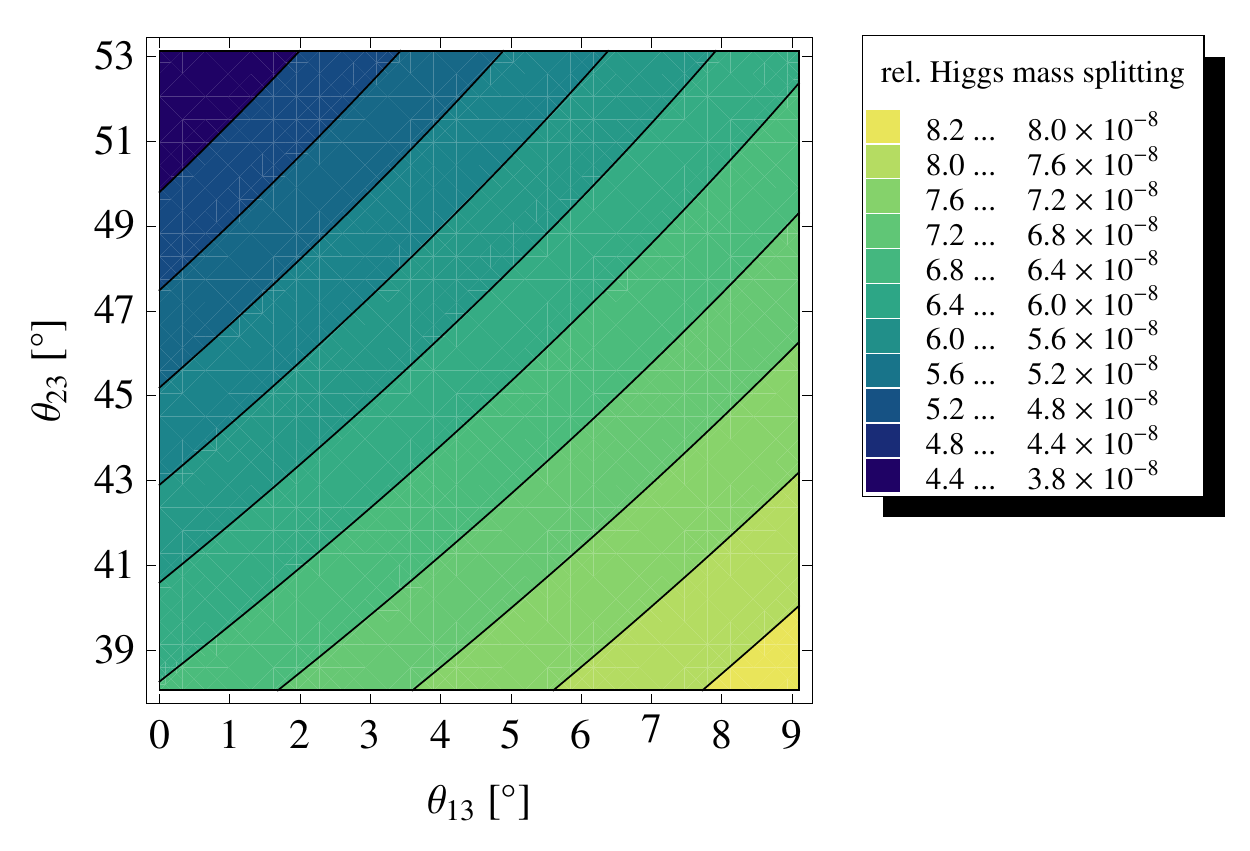}}
\caption{Correlations between $\theta_{13}$ and $\delta$ (a), $\theta_{23}$ and $\delta$ (b), and
$\theta_{13}$ and $\theta_{23}$ (c), for the matrix $P^L$ in (\ref{eq:PD54}). In (a),(b), and (c), we have set $\theta_{23}=45.0^\circ$, $\theta_{13}=9.1^\circ$, and $\delta=2/3\pi$, respectively. }\label{fig:D54}
\end{center}
\end{figure}

As a final example, consider the matrix representation
\begin{equation}\label{eq:PD54}
P^L=
\left(\begin{matrix}
0 & \omega & 0\\
1 & 0 & 0\\
0 & 0 & \omega^2
\end{matrix}
\right)
\end{equation}
for an element taken from the class $9\,C_3^{(1)}$ \cite{Escobar:2008vc} of $\Delta(54)$, which generates a $Z_6$ symmetry. In Fig.~(\ref{fig:D54}), we have summarized the resulting correlations between $\theta_{13}$, $\theta_{23}$, and $\delta$, for this $P^L$. The Higgs mass(-squared) splitting is defined as for the other examples. We see that the correlations between the leptonic mixing parameters differ strongly from those shown in  the other figures.

\end{document}